\documentclass[fleqn,usenatbib]{mnras}

\usepackage{newtxtext,newtxmath}

\usepackage[T1]{fontenc}

\DeclareRobustCommand{\VAN}[3]{#2}
\let\VANthebibliography\thebibliography
\def\thebibliography{\DeclareRobustCommand{\VAN}[3]{##3}\VANthebibliography}

\usepackage{adjustbox}  
\usepackage{amsmath}	
\usepackage{wasysym}    
\usepackage{bm}         
\usepackage{xspace}
\makeatletter 
  \patchcmd{\NAT@citex}
    {\@citea\NAT@hyper@{%
      \NAT@nmfmt{\NAT@nm}%
      \hyper@natlinkbreak{\NAT@aysep\NAT@spacechar}{\@citeb\@extra@b@citeb}%
      \NAT@date}}
    {\@citea\NAT@nmfmt{\NAT@nm}%
    \NAT@aysep\NAT@spacechar\NAT@hyper@{\NAT@date}}{}{}

  \patchcmd{\NAT@citex}
    {\@citea\NAT@hyper@{%
      \NAT@nmfmt{\NAT@nm}%
      \hyper@natlinkbreak{\NAT@spacechar\NAT@@open\if*#1*\else#1\NAT@spacechar\fi}%
        {\@citeb\@extra@b@citeb}%
      \NAT@date}}
    {\@citea\NAT@nmfmt{\NAT@nm}%
    \NAT@spacechar\NAT@@open\if*#1*\else#1\NAT@spacechar\fi\NAT@hyper@{\NAT@date}}
    {}{}
\makeatother

\newcommand\Msun{\text{M}_{\astrosun}} 
\newcommand\HI{\ion{H}{I}\xspace} 
\newcommand\HII{\ion{H}{II}\xspace} 
\newcommand\thesan{\mbox{\textsc{thesan}}\xspace}
\newcommand\thesanone{\mbox{\textsc{thesan-1}}\xspace}
\newcommand\thesantwo{\mbox{\textsc{thesan-2}}\xspace}
\newcommand\thesanwc{\mbox{\textsc{thesan-wc-2}}\xspace}
\newcommand\thesanhigh{\mbox{\textsc{thesan-high-2}}\xspace}
\newcommand\thesanlow{\mbox{\textsc{thesan-low-2}}\xspace}
\newcommand\thesansdao{\mbox{\textsc{thesan-sdao-2}}\xspace}

\newcommand\areport{\mbox{\textsc{arepo-rt}}\xspace}

\newcommand\zreion{$z_{\text{reion}}$\xspace}

\newcommand\orcid[1]{\href{http://orcid.org/#1}{\adjustbox{trim={-.15\width} {0\height} {-.15\width} {0\height},clip}{\includegraphics[height=12pt]{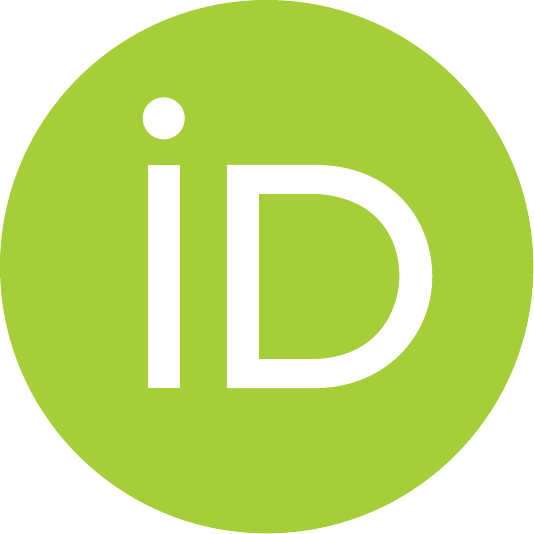}}}}

\title[Merging ionized bubbles in the EoR]{The \thesan project: tracking the expansion and merger histories of ionized bubbles during the Epoch of Reionization}

\author[N. Jamieson et al.]{%
Nathan~Jamieson\orcid{0009-0002-9160-6237},$^{1}$\thanks{E-mail: \href{mailto:njamieso@nd.edu}{njamieso@nd.edu}}
Aaron~Smith\orcid{0000-0002-2838-9033},$^{2}$
Meredith~Neyer\orcid{0000-0002-9205-9717},$^{3}$
Rahul~Kannan\orcid{0000-0001-6092-2187},$^{4}$
Enrico~Garaldi\orcid{0000-0002-6021-7020},$^{5}$
\newauthor
Mark~Vogelsberger\orcid{0000-0001-8593-7692},$^{3,6}$
Lars~Hernquist,$^{7}$
Oliver~Zier\orcid{0000-0003-1811-8915},$^{7}$
Xuejian~Shen\orcid{0000-0002-6196-823X}$^{3}$
and
Koki~Kakiichi\orcid{0000-0001-6874-1321}$^{8}$
\vspace{.1cm}%
\\%
$^{1}$Department of Physics and Astronomy, Notre Dame, Notre Dame, IN 46556, USA \\%
$^{2}$Department of Physics, The University of Texas at Dallas, Richardson, Texas 75080, USA \\%
$^{3}$Department of Physics $\&$ Kavli Institute for Astrophysics and Space Research, Massachusetts Institute of Technology, Cambridge, MA 02139, USA \\%
$^{4}$Department of Physics and Astronomy, York University, 4700 Keele Street, Toronto, ON M3J 1P3, Canada \\%
$^{5}$Institute for Fundamental Physics of the Universe, via Beirut 2, 34151 Trieste, Italy \\%
$^{6}$The NSF AI Institute for Artificial Intelligence and Fundamental Interactions, Massachusetts Institute of Technology, Cambridge MA 02139, USA \\%
$^{7}$Center for Astrophysics $\vert$ Harvard $\&$ Smithsonian, 60 Garden Street, Cambridge, MA 02138, USA \\%
$^{8}$Cosmic Dawn Center (DAWN), Niels Bohr Institute, University of Copenhagen, Jagtvej 128, DK-2200 Copenhagen N, Denmark%
}

\date{Accepted XXX. Received YYY; in original form ZZZ}

\pubyear{\the\year{}}

\begin{document}
\label{firstpage}
\pagerange{\pageref{firstpage}--\pageref{lastpage}}
\maketitle

\begin{abstract}
The growth of ionized hydrogen bubbles in the intergalactic medium around early luminous objects is a fundamental process
during the Epoch of Reionization (EoR). Observations using Lyman-$\alpha$ emission from high-redshift galaxies and forthcoming 21\,cm maps are beginning to constrain the sizes of these ionized regions. In this study, we analyze bubble sizes and their evolution using the state-of-the-art \thesan radiation-hydrodynamics simulation suite, which self-consistently models radiation transport and realistic galaxy formation throughout a large $(95.5\,\text{cMpc})^3$ volume of the Universe. Analogous to the accretion and merger tree histories employed in galaxy formation simulations, we characterize the growth and merger rates of ionized bubbles by focusing on the spatially-resolved redshift of reionization. By tracing the chronological expansion of bubbles, we partition the simulation volume and construct a natural ionization history. We identify three distinct stages of ionized bubble growth: (1) initial slow expansion around the earliest ionizing sources seeding formation sites, (2) accelerated growth through percolation as bubbles begin to merge, and (3) rapid expansion dominated by the largest bubble. Notably, we find that the largest bubble emerges by $z \approx 9\!-\!10$, well before the midpoint of reionization. This bubble becomes dominant during the second growth stage, and defines the third stage by rapidly expanding to eventually encompass the remainder of the simulation volume and becoming one of the few bubbles actively growing. Additionally, we observe a sharp decline in the number of bubbles with radii around $\sim 10$\,cMpc compared to smaller sizes, indicating a characteristic scale in the final segmented bubble size distribution. Overall, these chronologically sequenced spatial reconstructions offer crucial insights into the physical mechanisms driving ionized bubble growth during the EoR and provide a framework for interpreting the structure and evolution of reionization itself.
\end{abstract}




\section{Introduction}
As the first luminous objects formed and emitted hydrogen ionizing radiation billions of years ago, they initiated the Epoch of Reionization (EoR), spanning redshifts from $z\approx5\!-\!20$ \citep[EoR;][]{Shapiro1987, BarkanaLoeb2001, Furlanetto2006b, Wise2019}. The radiation from these early sources locally and inhomogeneously ionized the surrounding neutral hydrogen gas, creating ionized regions or ``bubbles'' in the intergalactic medium (IGM). Over time, these ionized bubbles grew, at first in isolation but eventually different bubbles began to overlap with each other, leading to percolation and a rapid increase in bubble sizes as they merged \citep{Furlanetto2016}. This process ultimately resulted in the nearly fully ionized Universe observed today. The EoR presents several astrophysical and cosmological frontiers, including understanding the physical mechanisms driving ionized bubble growth.

Observing galaxies, active galactic nuclei, and the ambient IGM during the EoR is challenging due to the extremely high redshifts. However, characterizing and interpreting observational signatures is of crucial importance, especially as recent and forthcoming \textit{James Webb Space Telescope} (\textit{JWST}) surveys and 21\,cm radio telescopes offer promising avenues to probe high-redshift galaxies and the properties of ionized bubbles \citep{Robertson2022}. To fully understand the EoR, studying both the IGM and galaxy sources is paramount, motivating investigations into the dynamics and morphologies of ionized bubbles and their connection to the local and large-scale environments \citep{Gnedin2022}.

The 21\,cm radio interferometers, including the Low Frequency Array \citep[LOFAR;][]{vanHaarlem2013}, Hydrogen Epoch of Reionization Array \citep[HERA;][]{DeBoer2017, HERA2023}, Square Kilometer Array \citep[SKA;][]{Mellema2013}, and others are beginning to map the distribution of neutral hydrogen in the Universe. The 21\,cm line, arising from the forbidden spin-flip hyperfine transition of neutral hydrogen, directly probes the IGM. By detecting the redshifted 21\,cm signal from the EoR, these instruments can reveal the global evolution, statistical spatial correlations, and even a tomographic picture of neutral and ionized gas. These measurements will be critical for studying ionized bubbles during the EoR as well as constraining cosmological parameters \citep{McQuinn2006, Mesinger2011, Liu2016, Park2019, Kannan2022b}.

In anticipation of the forthcoming 21\,cm data, numerous theoretical studies have sought to characterize the properties of ionized bubbles and their connections to the dominant processes in cosmic reionization. Different reionization scenarios and source models can lead to varying bubble morphologies; thus, the distribution of bubble sizes has emerged as a crucial diagnostic for distinguishing between reionization models \citep{McQuinn2007a, McQuinn2007b, Majumdar2016}.

Several methods have been developed to detect and analyze ionized bubble sizes within hydrodynamical simulations and semi-analytic models. Prominent among these are:
\vspace{-\topsep}
\begin{itemize}
  \item \textit{Mean-Free Path (MFP) Method:} Calculates effective bubble sizes by tracing rays from ionized cells to the nearest neutral cell in all directions \citep{Mesinger2007}.
  \item \textit{Spherical Averaging (SPA):} Determines the largest sphere over which the average ionization fraction exceeds a set ionization threshold \citep{Zahn2007}.
  \item \textit{Friends-of-Friends (FOF) Algorithm:} Links ionized cells within a given distance to form bubbles \citep{Ivezic2014}.
  \item \textit{Granulometry:} Uses a ``sieving'' process to count objects that fit within a given structural hierarchy \citep{Kakiichi2017}.
  \item \textit{Watershed Method:} Identifies bubbles by filling ``catchment basins'' from local minima until neighbouring basins intersect \citep{Lin2016}.
\end{itemize}
\vspace{-\topsep}

Beyond this, topological analyses have been fruitful in characterizing further distinctive signatures of reionization \citep{Friedrich2011, Busch2020, Giri2020JOSS, Elbers2023}.

Building upon these existing studies, we perform an analysis of ionized bubble sizes within the \thesan simulation suite \citep{Garaldi2022, Kannan2022a, Smith2022, Garaldi2023}. \thesan combines the galaxy formation model of IllustrisTNG \citep{Weinberger2017, Pillepich2018a, Pillepich2018b} with detailed on-the-fly modeling of radiative processes, including radiation transport \citep[\areport; ][]{Kannan2019}, non-equilibrium heating and cooling, and realistic ionizing sources, within a large cosmological volume of 95.5\,cMpc per box side. The \thesan simulations have been employed to make a wide range of EoR predictions \citep[e.g.][]{Kannan2022b, Qin2022, Borrow2023, Kannan2023, Yeh2023, Xu2023, Shen2024a, Shen2024b}.

Specifically, we examine the ionized bubbles by analyzing the reionization redshift (\zreion), which we define as the last time the ionized hydrogen fraction (\HII) crosses the threshold value of 0.5 from below \citep[e.g.][]{Thelie2022}. \zreion serves as an effective tracer for bubble sizes, particularly at scales below approximately 1\,cMpc \citep{meredith}. A distinction of our study is that we employ a chronological time-ordering of \zreion to construct a natural segmentation of the simulation volume. The resulting expansion and merger histories provide a unique and intuitive perspective on the evolution of ionized bubbles during the EoR.

The paper is organized as follows. In Section~\ref{sec:methods}, we describe our methods, including a brief overview of the \thesan simulation suite and the bubble tree algorithm used to analyze \zreion. In Section~\ref{sec:results}, we present our main findings, including the growth of ionized bubbles (\ref{subsec:growth_merge}) and the distribution of bubble sizes (\ref{subsec:bubblesize}) throughout the EoR. We explore the physical effects of different model parameters in Section~\ref{sec:physics_comparison}. Finally, we synthesize our conclusions in Section~\ref{sec:conclusion}. Supplementary discussion of grid resolution is found in Appendix~\ref{Appendix:Resolution Comparison} and the impact of corners being counted as neighbours in Appendix~\ref{Appendix:Corners as Neighbours}. All further mentions of ``reionization'' refer specifically to hydrogen reionization, and any mention of ``bubbles,'' ``bubble groups,'' or ``groups'' refers to ionized bubbles.

\section{Simulation and methods}
\label{sec:methods}
In this section, we provide an overview of the \thesan reionization simulations (\ref{sec:thesan}), present visualizations of the ionized bubbles (\ref{subsec:bubblevisualization}), describe in detail the algorithm used to analyze the ionized bubbles (\ref{sec:algorithm}), and discuss the resulting data products (\ref{sec:Algorithmdataprods}).

\begin{figure*}
    \centering
    \includegraphics[width=\textwidth]{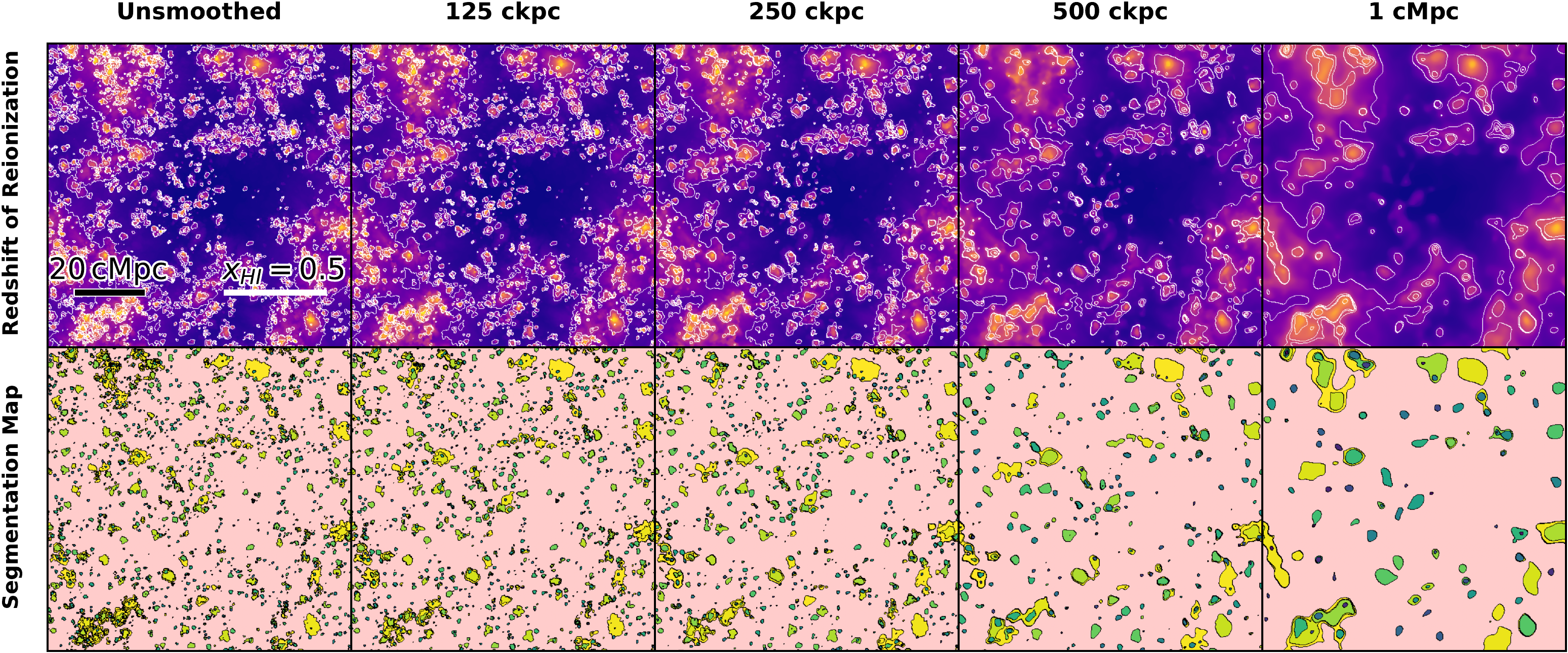}
    \caption{Visualizations of ionized bubbles displayed as cross-sections of the simulation volume. \textit{Top panels:} Slice images coloured according to the reionization redshift \zreion, with brighter colours indicating earlier reionization. White contours show concurrent \zreion fronts at $x_\HII=0.5$. \textit{Bottom panels:} The final segmented bubble groups at the end of the simulation ($z = 5.5$), represented by a different colour and outlined in black. The pink background is the largest group.}
    \label{fig:renderings}
\end{figure*}

\subsection{\thesan simulations}
\label{sec:thesan}
The \thesan project \citep{Kannan2022a, Garaldi2022, Smith2022, Garaldi2023} is a suite of large-volume ($L_{\text{box}} = 95.5\,\text{cMpc}$) radiation-magneto-hydrodynamic (RMHD) simulations designed to simultaneously resolve large-scale structures relevant for reionization and the detailed physics of realistic galaxy formation. It employs the IllustrisTNG galaxy formation model \citep{Pillepich2018a}, which is an updated version of the model used in the original Illustris simulation \citep{Vogelsberger2013, Vogelsberger2014b, Vogelsberger2014a}. \thesan offers state-of-the-art resolution and physics for reionization simulations at this volume, providing a unique opportunity to investigate connections between the topology of reionization and the galaxies responsible for driving this process.

Photons from various sources, including stars and active galactic nuclei, are tracked self-consistently in three energy bins: $13.6\,\text{eV}\!-\!24.6\,\text{eV}$, $24.6\,\text{eV}\!-\!54.4\,\text{eV}$, and $\geq 54.4\,\text{eV}$. Stellar population properties, such as luminosities and spectral energy distributions (SEDs), are determined using the Binary Population and Spectral Synthesis library \citep[BPASS;][]{Eldridge2017}. The simulations also incorporate non-equilibrium thermochemistry to track cooling by hydrogen and helium, as well as equilibrium cooling by metals.

The high-resolution, fiducial simulation, \thesanone, has dark matter and baryonic mass resolutions of $3.1 \times 10^6\, \rm \Msun$ and $5.8 \times 10^5\, \Msun$, respectively. Haloes are resolved down to masses of $M_\text{halo} \sim 10^8\, h^{-1}\, \Msun$ \citep{Garaldi2023}. The simulations use the efficient quasi-Lagrangian code \areport \citep{Kannan2019}, an extension of the moving mesh code \mbox{\textsc{arepo}}\xspace \citep{Springel2010, Weinberger2020} that includes radiative transport. Fluid dynamics equations are solved on an unstructured Voronoi mesh that adapts to the flow of the gas. Radiative transfer is handled using a moment-based approach with the M1 closure approximation \citep{Levermore1984, Dubroca1999}. Gravity is solved using a hybrid Tree-PM method, where short-range forces are computed using a hierarchical oct-tree algorithm \citep{BarnesHut1986}, and long-range potentials are calculated by solving the Poisson equation with Fourier methods.

The \thesan simulations output 81 snapshots of particle positions and properties, covering redshifts from $z = 20$ to $z = 5.5$. These particle snapshots are converted into Cartesian grids with varying resolutions: 128, 256, 512, and 1024 cells per side. These Cartesian grid representations encapsulate volume-weighted properties, including the ionized hydrogen fraction. In this paper, we utilize the grids with 512 cells per side. A discussion of the effects of different grid resolutions is provided in Appendix~\ref{Appendix:Resolution Comparison}.

Large-scale environmental properties such as the dark matter overdensity, $\delta \equiv (\rho - \bar{\rho}) / \bar{\rho}$ where $\bar{\rho}$ is the mean density, are most meaningful after smoothing over a given filter scale. Since bubble statistics are sensitive to small-scale features, it is advantageous to smooth out potentially transient or local fluctuations. Therefore, we also construct smoothed versions of the Cartesian outputs by first binning the particles at $1024^3$ resolution and then applying a periodic, mass-conserving Gaussian smoothing kernel. The smoothing scales are defined by standard deviations of 0\,ckpc (no smoothing), 125\,ckpc, 250\,ckpc, 500\,ckpc, and 1\,cMpc. We emphasize that the lower-resolution grids are coarse-grained versions of the high-resolution one ($1024^3$), with volume and mass weights correctly propagated.

\subsection{Visualizations of ionized bubbles}
\label{subsec:bubblevisualization}
Before diving into the details of our algorithm, we present visualizations of the ionized bubbles to provide context. Figure~\ref{fig:renderings} displays two different cross-sections of the ionized bubbles for each smoothing level. In the top panels, the slices are coloured according to the reionization redshift, \zreion, with brighter colours corresponding to higher \zreion values. The contours delineate the edges of concurrent ionized bubbles, shown when the global neutral hydrogen fraction equals $x_\HI = 0.5$. In the bottom panels, each unique bubble is depicted in a different colour and outlined in black, with the largest bubble highlighted in pink. These panels represent the state of the bubbles at the end of the simulation, at redshift $z=5.5$. There is only one bubble at $z=5.5$. However, the segmentation map freezes the earlier branches of the merger tree in their final state. These branches are represented with different colours.  

\begin{figure*}
    \centering
    \includegraphics[width=\textwidth]{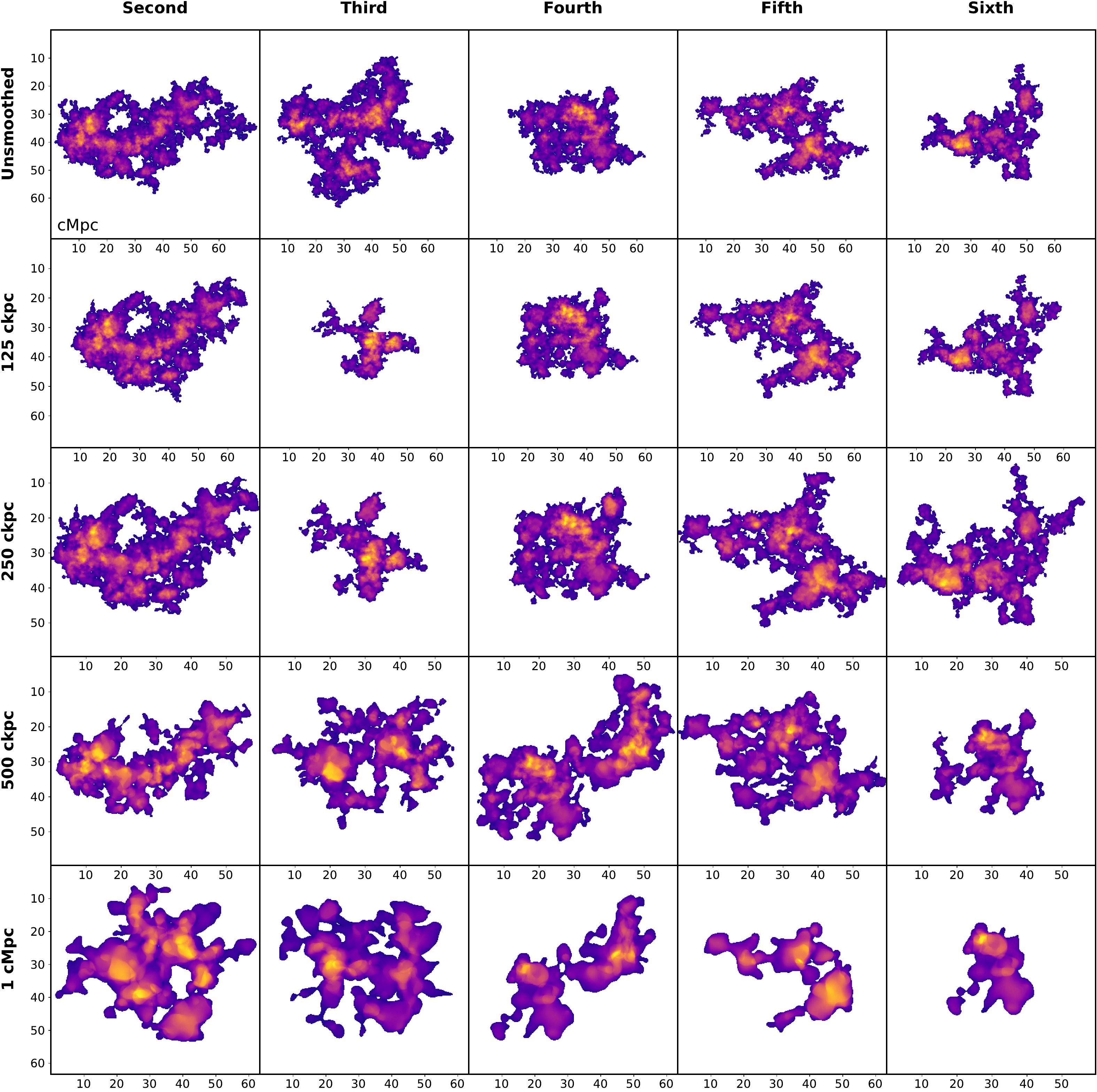}
    \caption{Line-of-sight projections of the second through sixth largest ionized bubbles for all smoothing levels (excluding the largest bubble). Each panel shows the projection along the $z$-axis, with brighter colours indicating a higher density of simulation cells belonging to each bubble. The images highlight the spatial distribution and morphology of the largest ionized regions prior to merging with larger bubbles.}
    \label{fig:projections}
\end{figure*}

To convey the variety of bubble morphologies, as well as the impact of pre-smoothing, Figure~\ref{fig:projections} presents a projection of the five largest bubbles, excluding the main bubble as it encompasses the entire simulation box at late times. The bubbles are projected along the $z$-axis, with brighter colours indicating a higher density of points, i.e., more cells belonging to the bubble along the line of sight. These visualizations illustrate the structure, distribution, and relative sizes of large spatially and temporally connected ionized regions prior to merging into larger bubbles. We note that the smoothing can change the ranking of bubble volumes, so we have been careful to arrange the columns to match the unsmoothed object ordering when possible.

\subsection{Bubble tree algorithm}
\label{sec:algorithm}
To analyze the growth of ionized bubbles, we developed an algorithm that utilizes the reionization redshift (\zreion) data from the \thesan simulations, which have periodic boundary conditions. The reionization redshift in each grid cell is defined as the last time the local ionized hydrogen fraction ($x_\HII$) crosses the threshold value of 0.5 from below. Our algorithm is designed to track the evolution of ionized regions by identifying individual bubbles, monitoring their expansion, and recording merger events. We run on several grid resolutions and smoothing levels: resolutions of $32$, $64$, $128$, $256$, and $512$ cells per side, and smoothing scales of 0\,ckpc (unsmoothed), 125\,ckpc, 250\,ckpc, 500\,ckpc, and 1\,cMpc. In this paper, we focus on the smoothing levels applied to the $512^3$ resolution grids. A discussion of the effects of different resolutions is provided in Appendix~\ref{Appendix:Resolution Comparison}. Fig.~\ref{fig:algorithmgraphic} illustrates the main steps of the algorithm applied to a simplified fictional dataset.

\noindent \textbf{\textit{Algorithm Overview:}} Our algorithm proceeds with the following main steps: find all bubble groups (maxima) within the reionization dataset, update neighbouring cell expansion queues, expand the correct group to the next lowest \zreion cell, and merge bubble groups until only one group remains. As the final group eventually encompasses the entire simulation volume, any cells that remain are automatically grown into by the last group. This includes all cells that are neutral at the end of the simulation, as we define these cells as being ionized at the final redshift. 

\noindent \textbf{\textit{Terminology:}} We define the following key terms before diving into the algorithm further:
\vspace{-\topsep}
\begin{itemize}
  \item \textit{Value:} The reionization redshift at a given point.
  \item \textit{Point:} A coordinate cell in the 3D dataset.
  \item \textit{Adjacent Points:} The 26 neighbouring points surrounding a given point (including diagonals). Adjacent values correspond to the values of the adjacent points.
  \item \textit{Neighbouring points:} For an entire bubble group, all adjacent points that are not already part of the bubble group.
\end{itemize}
\vspace{-\topsep}
We compare the effects of either including or excluding corner neighbours (reducing from 26 to 6 adjacent points) in Appendix \ref{Appendix:Corners as Neighbours}.

\begin{figure*}
    \centering
    \includegraphics[width=\textwidth]{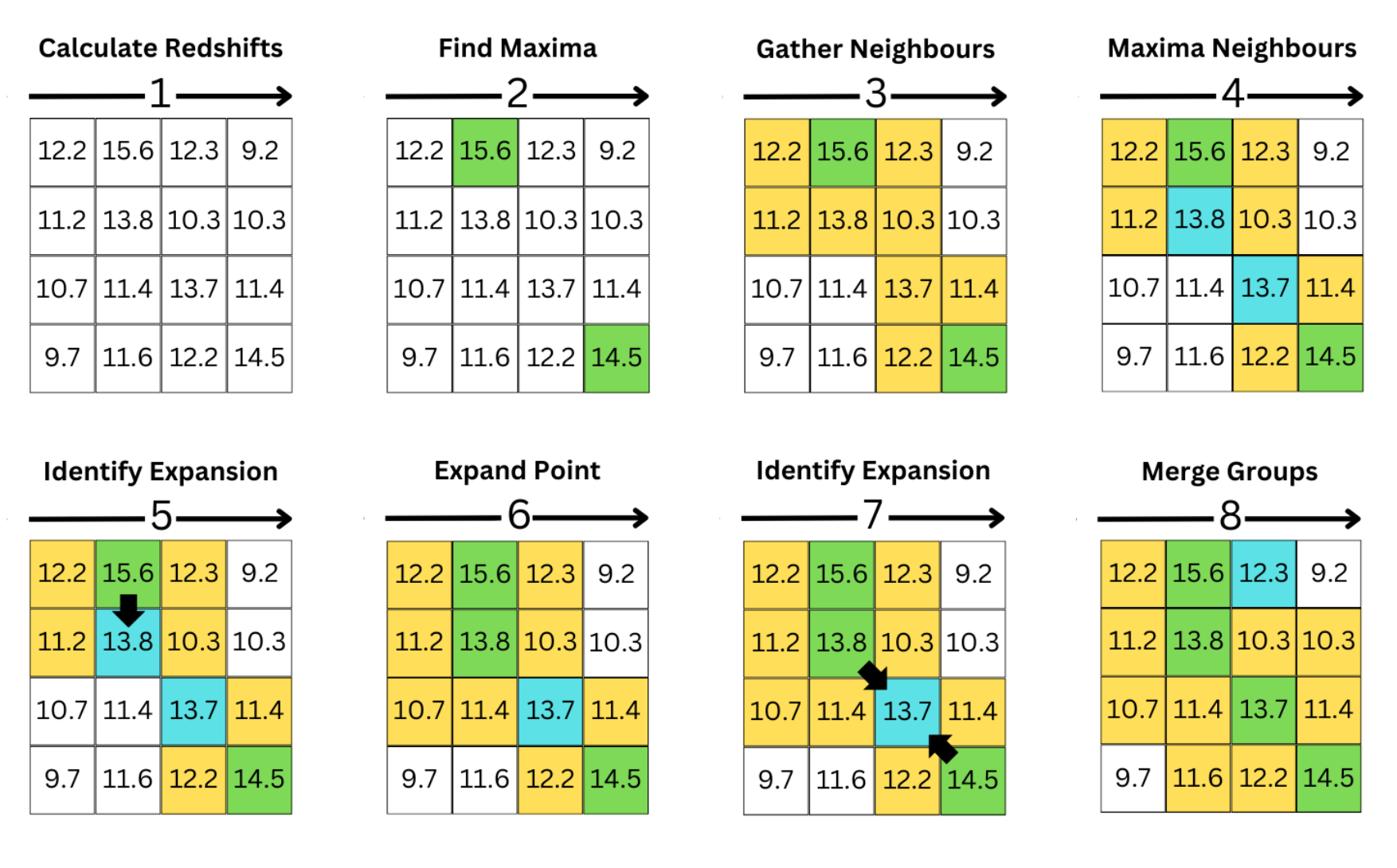}
    \caption{Illustration of the bubble tree algorithm applied to a fictional dataset (assuming non-periodic boundary conditions). \textit{Panel 1:} The algorithm is applied to the redshift of reionization \zreion field. \textit{Panel 2:} Local maxima are identified (highlighted in green) as initial bubble groups. \textit{Panel 3:} Neighbouring points (highlighted in yellow) are collected for each bubble. \textit{Panel 4:} The highest neighbouring values (highlighted in blue) are identified for each bubble. \textit{Panel 5:} The bubble with the highest neighbouring value is selected for expansion (indicated by the arrow). \textit{Panel 6:} The selected bubble expands into the neighbouring point, updates its next neighbouring points, and updates its expansion queue of the next highest neighbouring values. \textit{Panel 7:} Two bubbles identify the same highest neighbouring value as their next expansion point. \textit{Panel 8:} The two bubble groups merge into a single larger group and continue expanding as one.}
    \label{fig:algorithmgraphic}
\end{figure*}

\noindent \textbf{\textit{Initial Setup:}} The algorithm begins by identifying all local maxima in the \zreion dataset, which serve as the initial seeds of bubble groups. A local maximum is defined as a point whose \zreion value is greater than all of its adjacent neighbour values. If an adjacent value is equal to the value of the current point, it is not considered a true local maximum and is not initialized as a separate bubble.

\noindent \textbf{\textit{Expansion Queue Creation:}} Once the bubbles are identified, for each group, we create a queue for the next expansion points by collecting all neighbouring points and sorting them in decreasing order of their \zreion values. The first element in this sorted list represents the highest neighbouring value and serves as the next potential expansion point for the bubble.

\noindent \textbf{\textit{Expansion and Merging:}} With the queues initialized, the algorithm then iteratively expands the bubble groups into the neighbouring points with the highest \zreion values. Within each iteration loop, the algorithm first identifies all bubble groups that have the same highest neighbouring \zreion value, which signals one or more potential mergers. To facilitate this search, we maintain a ranked list of the maxima of the neighbouring values of each bubble group. For each of these bubble groups, it also checks if they have any additional equivalent neighbouring values. Once all points are found, they are expanded. If no mergers are detected, active bubble groups expand into their highest neighbouring point. The new points are added to their corresponding bubble group and removed from the group's neighbouring point list. Next, it finds the neighbouring values of the newly expanded points and inserts them into the group's list if it is not already a neighbour or a point in the group. Lastly, it updates the group's expansion queue for each new value.

A merger event is triggered whenever two or more groups attempt to expand into the same point, which is recorded when one group attempts to add a point belonging to another group. In the case that the neighbouring point is shared by multiple groups on the same iteration, only one group will freely move into the point while the other groups that move into the point will trigger their own merger events. The algorithm proceeds with normal expansion for all groups that are not flagged as potential mergers. Furthermore, no data gets stored that establishes any group as moving into contested points until the merge events are resolved.

To resolve a series of mergers, we first gather all bubble groups involved. Each unique neighbouring point that triggers a merging event represents two or more bubble groups. From the merging groups, we identify the largest bubble group (by volume) as the parent group and merge the smaller groups into it. The parent group expands into the neighbouring point of each merging group and updates its bubble size and other characteristics to include the merged groups. The neighbouring points of the smaller groups are inserted into the parent group's neighbouring points if they are not already included. Finally, the expansion queue of the largest group is updated to include all the new neighbouring values, and the smaller groups are removed from further consideration as they no longer expand independently.

After resolving the highest neighbouring value and the corresponding neighbouring points, the algorithm repeats the expansion and merging steps until only one bubble group remains. Once this occurs, any remaining points are added to the last bubble group, effectively considering them as uncontested growth ionized between the final merger event and the final redshift. 

\noindent \textbf{\textit{Comparison with the Watershed Method:}} At first glance, our proposed method may appear identical to the watershed algorithm \citep[e.g.][]{Lin2016}, but there are key differences beyond simply applying it to the \zreion field. In the watershed method, regions are grown from local minima upward, and merging occurs when basins meet. Importantly, in the watershed method, regions continue to grow even after merging with others, so the boundaries of the resulting volumes represent points of first contact. In contrast, our algorithm constructs a series of merging bubble groups. Once a bubble merges with a larger one, it stops growing independently. This approach leads to bubbles of different sizes and shapes, reflecting the hierarchical nature of bubble growth during reionization mirroring the large-scale structure formation of the Universe. 

\subsection{Bubble tree code and data products}
\label{sec:Algorithmdataprods}
A version of our C++ bubble tree code is found at \url{https://github.com/NJamieson22/Bubble-Merger-Tree}.
After running the algorithm, we store the resulting data products in an HDF5 file with the following naming convention: \verb"[smoothing]_[resolution]_tree_data.hdf5", where the terms in square brackets encoding the smoothing level and resolution of the run, respectively. The HDF5 file contains several datasets that record various properties of the bubbles, their expansion, and merging history. The lengths are typically equal to the number of bubble groups, and the index represents the bubble group, sorted by decreasing total volume including all progenitors. A quick reference to the saved attributes is provided in Table~\ref{tab:saveddata}.

Key datasets for merger event statistics include:
\vspace{-\topsep}
\begin{itemize}
  \item \verb"merged_with": A 1D array indicating, for each bubble group, the group it merged with. Only the final group does not merge into a larger group.
  \item \verb"z_merge": The reionization redshift at which each bubble group merged with another one.
  \item \verb"z_form": The reionization redshift at which each bubble group formed, i.e., the \zreion of its initial local maximum.
  \item \verb"cells_merged" and \verb"parent_cells_merged": The number of cells in each bubble group at the time of merging, and the size of the parent group it merged into (including all progenitors).
  \item \verb"HII_Z_count": The global \HII fraction as the algorithm runs.
  \item \verb"i_com" and \verb"dr_com": The moment of inertia tensor and centre-of-mass coordinates for each bubble group at the time of merging. Specifically, \verb"dr_com" records the full $\{x, y, z\}$ centre-of-mass position relative to the initial cell defining the bubble group, and \verb"i_com" stores the flattened symmetric components with in the following order: $x^2$, $y^2$, $z^2$, $x \cdot y$, $x \cdot z$, and $y \cdot z$.
\end{itemize}

Key datasets for growth statistics include:
\vspace{-\topsep}
\begin{itemize}
  \item \verb"counts": The number of cells each bubble group expanded into through its own growth (excluding merged cells). The sum of this array is equal to the total number of grid cells.
  \item \verb"bubble_cells": An array of coordinates for all cells that each bubble group expanded into. The size is equal to three times the number of grid cells as it stores the $\{x, y, z\}$ values added during the expansion stage of the algorithm.
  \item \verb"offsets": An array used to index into \verb"bubble_cells" to retrieve the coordinates of each bubble group. The size is one more than the number of bubble groups for intuitive slice indexing.
  \item \verb"cell_to_bubble": A 3D array mapping each cell in the grid to the bubble group that grew into that cell.
  \item \verb"effective_volume": The effective volume of each point when a bubble group expands into that cell.
\end{itemize}

At the end of the algorithm, the data products are sorted such that the largest bubble group is listed first, the second largest second, and so on, based on the \verb"cells_merged" attribute. The last remaining bubble is considered to merge with itself at the final redshift.

\begin{table}
  \centering
  \caption{Summary of the data products saved by the bubble tree algorithm. For datasets with sizes equal to the number of bubble groups, the index corresponds to the bubble group ID, sorted in order of decreasing volume upon merging. See the text for further details.}
  \label{tab:saveddata}
  \begin{tabular}{ll} 
    \hline
    Name & Description \\
    \hline
    \verb"merged_with" & ID of the group each bubble merged with\\
    \verb"z_merge" & Redshift at which each bubble group merged\\
    \verb"z_form" & Redshift at which each bubble group formed\\
    \verb"cells_merged" & Number of cells in the merging bubble group\\
    \verb"parent_cells_merged" & Number of cells in the parent bubble group\\
    \verb"counts" & Number of expansion cells for each group\\
    \verb"bubble_cells" & Cell coordinates each bubble expanded into\\
    \verb"offsets" & Offsets to index into \verb"bubble_cells"\\
    \verb"HII_Z_count" & Number of ionized cells at each redshift\\
    \verb"effective_volume" & Effective volume of each cell at expansion\\
    \verb"cell_to_bubble" & Group that grew into the coordinate\\
    \verb"i_com" & Moment of inertia tensor for each bubble\\
    \verb"dr_com" & Relative centre-of-mass coordinates $\{x, y, z\}$\\
    \hline
  \end{tabular}
\end{table}

\section{Results}
\label{sec:results}
In this section, we present our analysis of tracking ionized bubbles during the EoR. We first analyze the growth characteristics of the bubbles in Section~\ref{subsec:growth_merge}, followed by a discussion of their size distribution and morphology in Section~\ref{subsec:bubblesize}.

\subsection{Bubble growth}
\label{subsec:growth_merge}
We divide our findings for bubble growth into three parts. In Section~\ref{subsubsec:global}, we discuss the global reionization history and the overall growth rate of ionized regions. In Section~\ref{subsubsec:presistentreion}, we examine the growth characteristics for persistent bubbles. Finally, in Section~\ref{subsubsec:largestbub}, we focus on the evolution of the largest ionized bubble group.

\begin{figure}
    \centering
    \includegraphics[width=\columnwidth]{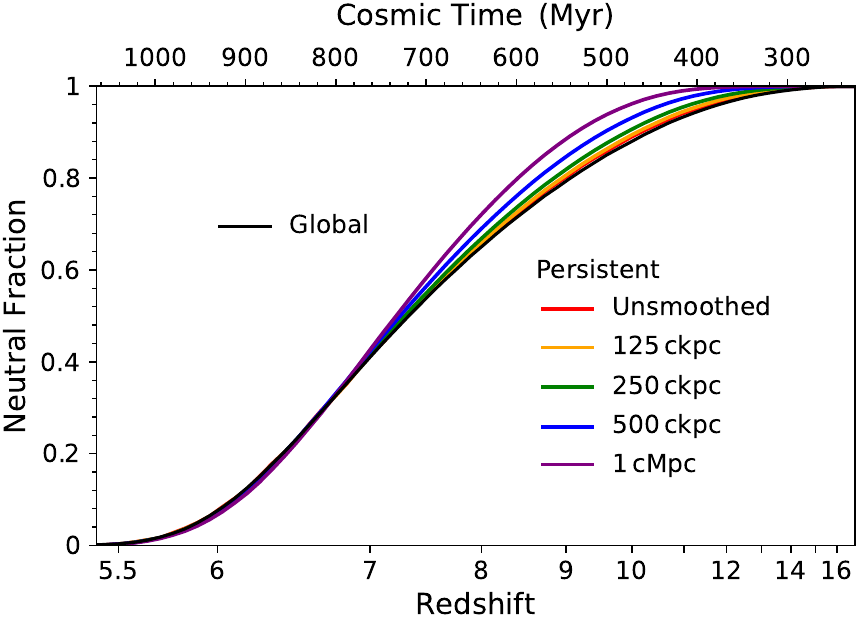}
    \caption{\textbf{Global Reionization History:} The volume-averaged neutral hydrogen fraction ($x_\HI$) as a function of redshift for different smoothing levels. The black curve represents the global $x_\HI$ from the \thesan simulation, while coloured curves correspond to the persistent \zreion-based histories with different smoothing scales. Higher smoothing levels deviate from the global history and exhibit larger $x_\HI$ at early times. The no smoothing and $125$\,ckpc cases are nearly identical to the true evolution; however, there is already a significant deviation observed with $500$\,ckpc.}
    \label{fig:HI}
\end{figure}

\subsubsection{Global reionization history and growth rate}
\label{subsubsec:global}
As luminous sources begin to ionize the surrounding neutral hydrogen gas, the volume-averaged fraction of ionized hydrogen ($x_{\HII} \equiv n_{\HII} / n_\text{H}$) increases while the neutral component decreases ($x_{\HI} = 1 - x_{\HII}$). These fractions allow us to construct the global reionization history as a function of redshift or cosmic time.
Figure~\ref{fig:HI} illustrates the global reionization history extracted from the \thesan simulation, along with the persistent reionization histories obtained from the \zreion data at different smoothing levels. The unsmoothed reionization history closely matches the global evolution, while higher smoothing levels (especially at 500\,ckpc) begin to deviate from the true history, i.e. $x_{\HI}$ is larger at early times compared to lower smoothing levels. This divergence occurs because increased smoothing suppresses small-scale fluctuations, leading to differences in the ionization topology. As a result, bubbles smaller than the smoothing scale are effectively infilled, impacting the reionization history more significantly at higher redshifts, while the histories converge at lower redshifts. The initially bursty star formation leads to many small bubbles that are not persistent in time, so smoothing both reduces the number of bubbles and delays the final redshift of reionization.

\begin{figure}
    \centering
    \includegraphics[width=\columnwidth]{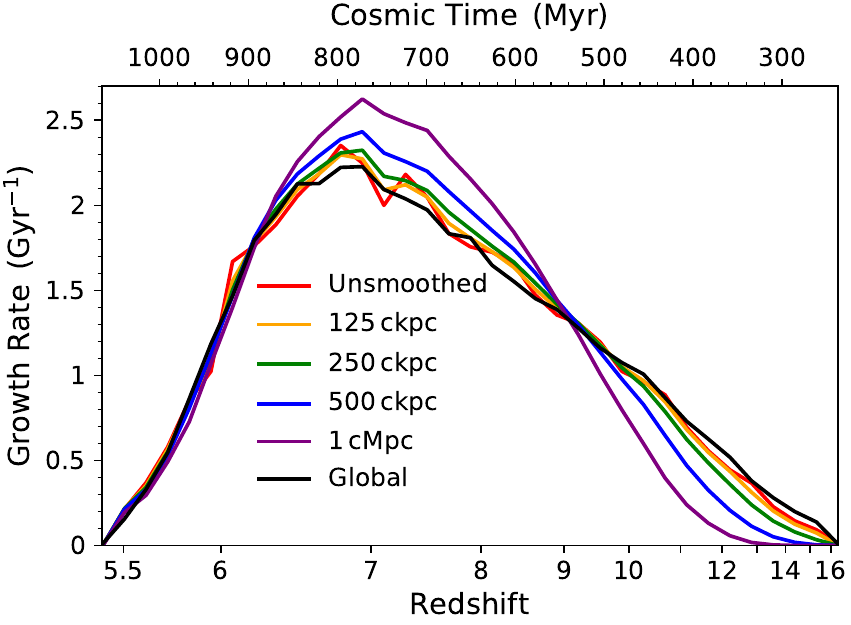}
    \caption{\textbf{Global Growth Rate:} The growth rate of ionized regions as a function of redshift for different smoothing levels, normalized by the simulation volume. Each bin represents a constant time interval of approximately $32\,$Myr. During the ramp up phase higher smoothing levels show lower growth rates at $z > 9$, and higher growth rates at $z < 9$. The differences in bubble expansion and interactions are due to the suppression of small bubbles at early times, delaying the reionization redshift \zreion as in Figure~\ref{fig:HI}.}
    \label{fig:growthrate}
\end{figure}

In Figure~\ref{fig:growthrate}, we show the growth rate of ionized regions as a function of redshift to assess the relative importance of each epoch as reionization progresses. We define the growth rate as the fraction of simulated volume that is ionized per unit time. The growth rate initially increases as the ionized bubbles expand and more cells become available for ionization. It then peaks around redshift $z \approx 7$ and gradually decreases as the number of neutral cells diminishes, leading to a natural decline in the growth rate. We can compare the global growth rate to the persistent growth rate in a similar manner as Figure~\ref{fig:HI}. As expected, lower smoothing levels are more closely aligned with the global growth rate. Furthermore, the graph reveals two areas of interest during the ramp-up phase. For higher redshifts ($z > 9$), the growth rate is suppressed as the smoothing scale increases compared to the instantaneous growth rate. At lower redshifts ($z < 9$), higher smoothing levels have enhanced growth rates to catch up after delayed growth, allowing larger ionized regions to expand more rapidly into the remaining neutral regions. The highest smoothing scale also exhibits the highest peak growth rate (at $z \approx 7$).

These phenomena also manifest in the bubble tree results and can be explained by the number of bubbles present in each redshift range. At lower redshifts, lower smoothing results in more bubbles, leading to more potential expansion points early on. Therefore, a higher expansion rate is expected, as confirmed by the figure. Additionally, the higher expansion rate for lower smoothing is consistent with the results from Figure~\ref{fig:HI}, as the \HI fraction increases with increased smoothing at low redshift, implying a higher growth rate for lower smoothing ($z > 9$). The higher smoothing levels accelerate to catch up to the converged evolution at $z < 7$ and ionize the rest of the box at the same time regardless of the smoothing scale. The critical redshift when the highest growth rates swap ($z\approx 9$), corresponds to the time when small bubbles begin to merge into larger ones. When they merge, the points the merging groups expanded into have already been ionized, so they do not count as new growth in the growth rate. Meanwhile at higher smoothing levels, the bubbles expand into unclaimed neutral regions at later times.

\begin{figure}
    \centering
    \includegraphics[width=\columnwidth]{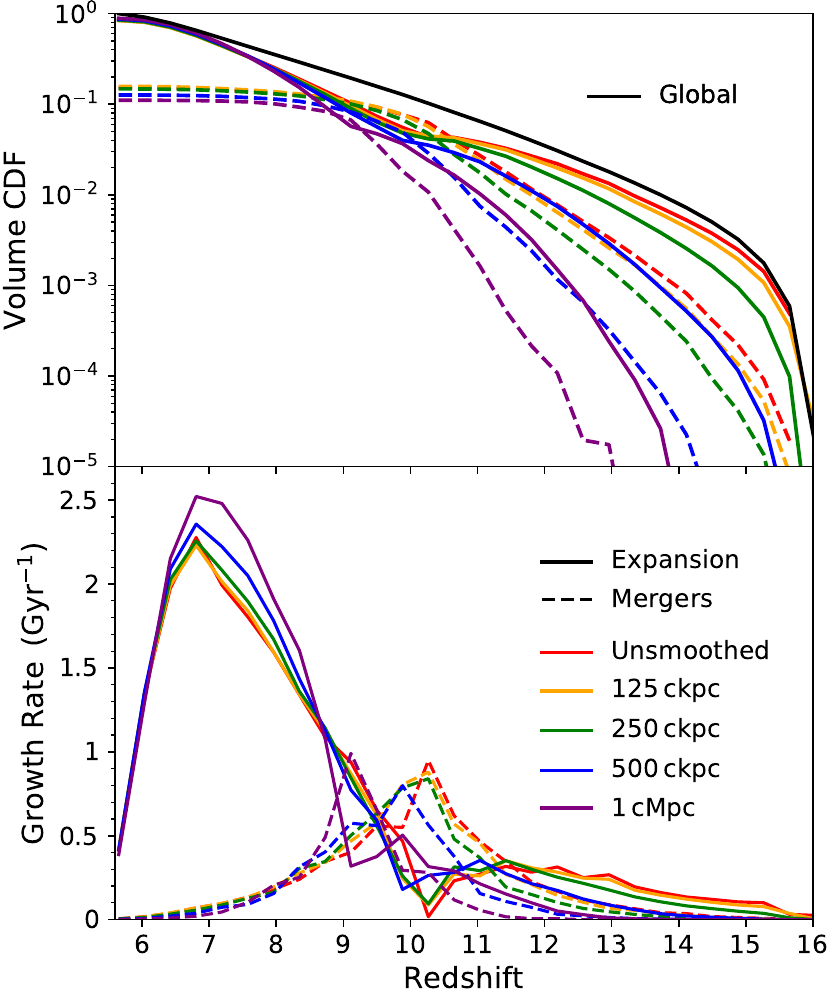}
    \caption{\textbf{Growth by Expansion and Mergers:} At any redshift, the \zreion data can be partitioned into bubbles that have merged, bubbles that have not merged but expanded up to that redshift, and points not yet ionized. \textit{Top panel:} Cumulative volume of cells added through expansion and merging for redshifts greater than or equal to $z$. \textit{Bottom panel:} The corresponding expansion and merging rates per unit volume and time interval. Three distinct growth phases are evident: initial expansion dominated by small bubbles, a peak in merging activity as bubbles begin to overlap, and a final phase of rapid expansion by the largest bubble. The ordering of smoothing levels reflects the number of bubble groups present at each stage.}
    \label{fig:prepost}
\end{figure}

\subsubsection{Persistent reionization history}
\label{subsubsec:presistentreion}
To gain deeper insight into the growth dynamics of ionized regions during the EoR, we analyze the expansion and merging behaviours of individual bubbles over time. Here, expansion refers to a bubble growing by ionizing neighbouring neutral cells that have not yet merged with any other bubbles. In contrast, merged points are cells that become part of a larger bubble through merging events. For instance, when bubble one merges into bubble two, all its expansion points are reclassified as merged points and are no longer considered part of independent expansion.

Figure~\ref{fig:prepost} presents a complementary view of the growth of ionized bubbles as a function of redshift. The top panel shows the cumulative ($\geq z$) volume added through expansion and merging. The bottom panel displays the corresponding expansion and merging rates per unit time, normalized by the simulation volume. Cells not yet ionized at a given redshift are excluded from these counts.

\begin{figure}
    \centering
    \includegraphics[width=\columnwidth]{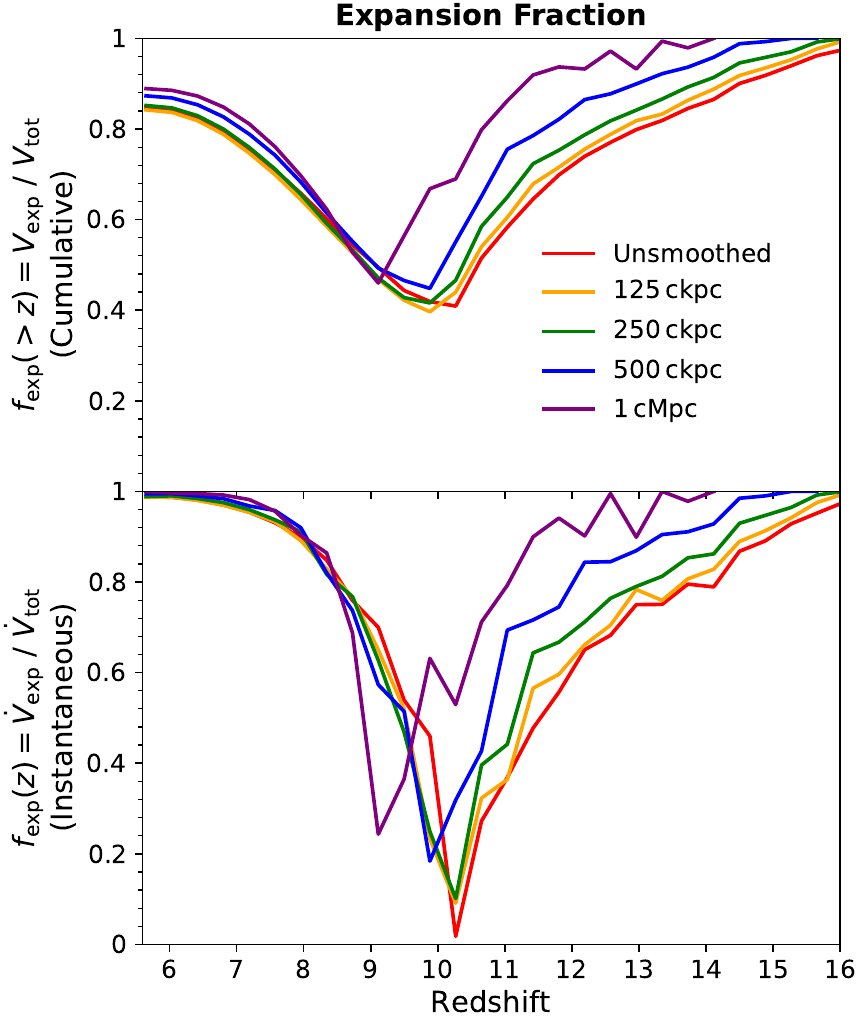}
    \caption{\textbf{Expansion Fraction During Growth Phases:} The expansion fraction highlights the relative contributions of expansion and merging throughout reionization. \textit{Top panel}: Cumulative expansion fraction above a given redshift, $f_\text{exp}(>z) = V_\text{exp} / V_\text{tot}$. \textit{Bottom panel}: Instantaneous expansion fraction at a given redshift, $f_\text{exp}(z) = \dot{V}_\text{exp} / \dot{V}_\text{tot}$. The three growth phases are evident, with the expansion fraction decreasing during the merging phase and increasing again during the final rapid expansion. The ordering of smoothing levels reflects the starting number of bubble groups and delayed \zreion, both impacting the timing of these transitions.}
    \label{fig:prepostfrac}
\end{figure}

The progression in Figure~\ref{fig:prepost} reveals three distinct stages in the evolution of ionized bubbles.
\vspace{-\topsep}
\begin{itemize}
  \item \textit{Initial Expansion Phase ($z \gtrsim 11$):} At the highest redshifts, bubbles start as single-cell regions corresponding to local maxima in \zreion. They grow by expanding into adjacent neutral cells, with minimal merging since they are widely separated. The expansion rate increases gradually as bubbles grow larger and encounter more neighboring cells to ionize.
  \item \textit{Merging Phase ($z \approx 9-11$):} As bubbles expand, they eventually come into contact with neighboring bubbles, leading to an increase in merging events. The merging rate spikes, as seen in the bottom panel of Figure~\ref{fig:prepost}, while the expansion rate dips as growth focuses on combining existing ionized regions rather than ionizing new cells.
  \item \textit{Rapid Expansion Phase ($z \lesssim 9$):} After the peak in merging activity, the largest bubble begins to dominate the simulation volume. It rapidly expands into the remaining neutral regions, causing a sharp increase in the expansion rate as the surface area of ionization fronts peaks. This acceleration continues until the supply of neutral cells is exhausted, naturally decreasing the expansion rate as reionization completes.
\end{itemize}
\vspace{-\topsep}

In regards to the differences between the smoothing levels, in the first stage, lower smoothing levels result in higher expansion rates because they preserve small-scale fluctuations, resulting in more initial bubbles. With more bubbles, there are more opportunities for expansion into neutral regions early on.
During the merging phase, the unsmoothed data exhibits the earliest rise and peak in merging activity, with this transition occurring later as the smoothing increases. With more bubble groups, merging starts earlier due to the increased likelihood of interaction between bubbles. The greater density of small-scale structures, and corresponding reduced mean bubble separation, produces more frequent mergers at earlier times. Higher smoothing levels also delay the onset of merging since fewer bubbles are present initially, allowing a longer free expansion into neutral regions without encountering other bubbles. Therefore, the peak of merging occurs at a lower redshift. However, runaway percolation leads to a rapid increase in the dominant bubble sizes, and a drastic reduction in the number of surviving isolated bubbles.
In the final phase of rapid expansion, higher smoothing levels exhibit higher expansion rates mainly to catch up from the delayed \zreion. In the late stages, the merging rates are not significantly different, which means the number of remaining bubbles is similar, erasing the earlier distinction between many small bubbles or fewer medium-sized ones. Interestingly, the difference in the final cumulative expansion and merger volumes is equal to the volume of the largest bubble group, which represents a vast majority of the simulated volume.

To further assess the balance between expansion and merging, we introduce the expansion fraction ($f_\text{exp}$), defined as the ratio of the number of cells added through expansion ($V_\text{exp}$) to the total number of cells added (i.e. expansion plus merging, ($V_\text{tot}$)). Figure~\ref{fig:prepostfrac} plots the expansion fraction both cumulatively (top panel) and per redshift bin (bottom panel). From this figure, we can confirm the results observed earlier about the three growth phases. In the initial expansion phase, $f_\text{exp}$ is close to unity, indicating that most of the growth is due to expansion, smoothly decreasing with time. During the merging phase, $f_\text{exp}$ drops sharper, especially for lower smoothing levels, signifying that growth is dominated by mergers. For a short time, the instantaneous $f_\text{exp}$ even approaches zero for the unsmoothed data, indicating that almost no new expansion occurs during this stage outside of merging. In the final rapid expansion phase, $f_\text{exp}$ rises again as the largest bubble expands into remaining neutral regions. Furthermore, the timing of these phases shifts with smoothing level. Lower smoothing levels experience earlier transitions due to the larger number of bubbles and increased interactions.

\begin{figure}
    \centering
    \includegraphics[width=\columnwidth]{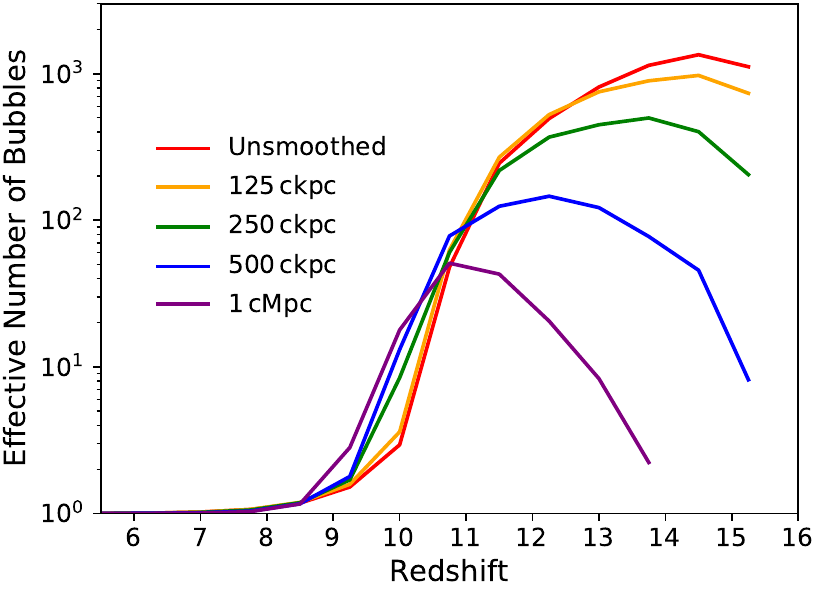}
    \caption{\textbf{Effective Number of Active Bubbles:} The effective number of actively growing bubbles as a function of redshift for different smoothing levels. Lower smoothing levels have higher numbers of contributing bubbles at early times, peaking earlier and at higher values. This occurs because higher smoothing generates fewer bubbles initially and is more affected by the damping of \zreion structure. At lower redshifts, the effective number converges to one across all smoothing levels, indicating the dominance of a single large bubble in the later stages of reionization.}
    \label{fig:neff}
\end{figure}

The last statistic we examine is the effective number of actively growing bubbles at each redshift. This quantity is defined as $n_\text{eff} \equiv \langle V \rangle^2 / \langle V^2 \rangle$, where $V$ denotes each bubble volume and the angle brackets are averages over all bubbles, and provides a quantitative measure of the relative importance of bubbles contributing to ionization. Figure~\ref{fig:neff} plots the effective number of bubbles for each smoothing level. The lowest smoothing level has the highest effective number of bubbles at high redshifts ($z \gtrsim 11$), which decreases as smoothing increases. This is expected, as lower smoothing results in more bubbles due to more small-scale structures being preserved. Second, the peak of the effective number of bubbles shifts to lower redshifts as smoothing increases. This occurs because smoothing dampens the reionization values, causing a slower initial growth. Lastly, at low redshifts ($z \lesssim 9$), the effective number of bubbles converges to one for all smoothing levels, indicating that a single bubble begins to dominate the growth. In summary, bubble growth dynamics provide insights into the phases of reionization, and are impacted by smoothing. However, the eventual dominance of the largest bubble is a consistent outcome.

\subsubsection{Growth of the largest bubble}
\label{subsubsec:largestbub}
\begin{figure}
    \centering
    \includegraphics[width=\columnwidth]{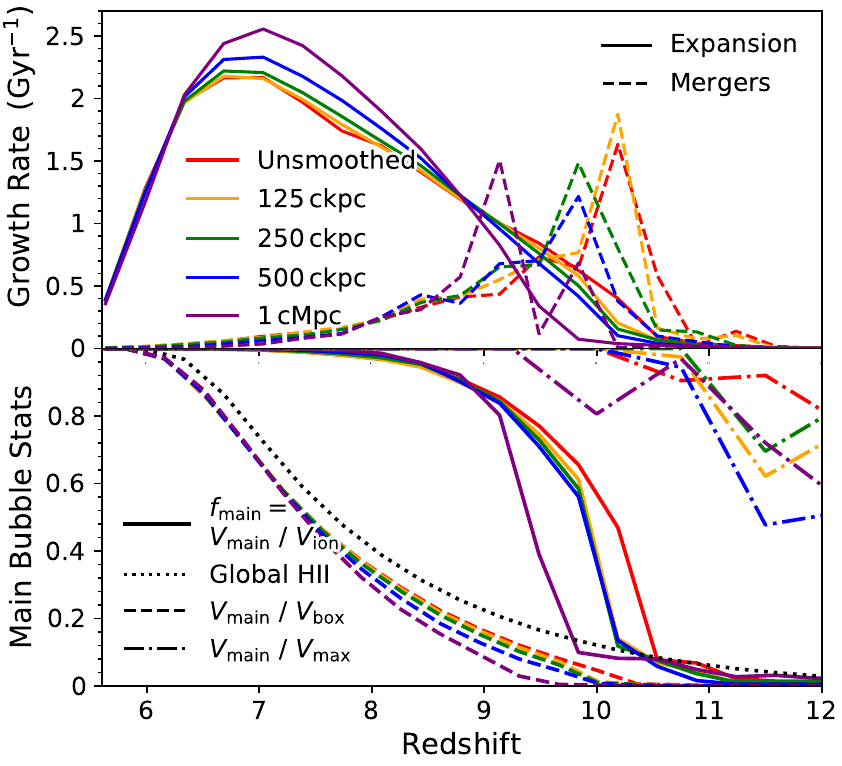}
    \caption{\textbf{Growth of the Largest Bubble:} Similar to the global growth from Figure~\ref{fig:prepost} but for the main bubble group. \textit{Top panel:} The expansion and merging growth rates of the largest bubble as a function of redshift. \textit{Bottom panel:} The fraction of ionized volume in the main bubble $f_\text{main} \equiv V_\text{main} / V_\text{ion}$, where $V_\text{ion}=\sum_i V_i$ (\zreion>z).
    The growth is relatively slow until it merges with other bubbles, where it rapidly increases the fraction. With the largest bubble and the box undergoing rapid expansion, the fraction increases to one. The largest bubble establishes itself after its main merging event, dominating the growth of the box thereafter. For comparison, defining $V_\text{box}$ as the total simulated volume and $V_\text{max}$ as the volume of the largest bubble at any given time, we also show both the fraction of the box this corresponds to, $V_\text{main} / V_\text{box}$, which mirrors the global ionized fraction, and the fraction of the current largest bubble, $V_\text{main} / V_\text{max}$, which becomes the main bubble at $z\sim 10.5$.}
    \label{fig:fmain}
\end{figure}

During the final expansion phase of reionization, the entire simulation volume becomes encompassed by a single bubble group, which we refer to as the largest or main bubble. We now apply the same expansion and merging metrics used to examine the entire box to this largest bubble. In the top panel of Figure~\ref{fig:fmain}, we show the growth rate history for the main bubble separated by expansion and merging. The expansion points are the regions the largest bubble expanded into, while the merged points are all of the bubble groups that merged into the largest bubble. In the bottom panel, we plot the fraction of the total ionized volume grown into by the largest bubble, $f_\text{main} \equiv V_\text{main} / \sum V_i$, emphasizing that this is compared to the cumulative persistent ionized volume in terms of \zreion at each redshift.

The growth rates for the largest eventual bubble (top panel of Figure~\ref{fig:fmain}) are similar to those from the entire simulated volume (bottom panel of Figure~\ref{fig:prepost}), including the three stages of growth: initial expansion, rapid mergers, and accelerated expansion. However, there are important differences between these stages. First, the initial expansion stage is much less important than in the full box. This is because the expansion around any given bubble starts off slowly, ionizing a few cells at a time, whereas in the globally-averaged scenario the early ramp up phase has contributions from a high density of these small bubbles. Therefore, there is nothing particularly special about the main bubble at these times. Second, the merging stage exhibits a higher peak and shorter width than in Figure~\ref{fig:prepost}. A smaller number of bubbles merge into the largest group than the total number of mergers in the simulation (as each bubble that merges into the largest group has its own merging history). Consequently, there are relatively more cells in each of the bubbles that merge into the largest bubble and fewer bubbles directly involved in the merging process. This results in a higher peak and shorter width for the merging stage. In fact, the distinguishing feature in this phase is that the main bubble is always the primary (largest volume bubble group) in every merger event, while secondaries behave like branches with truncated histories.

\begin{figure}
    \centering
    \includegraphics[width=\columnwidth]{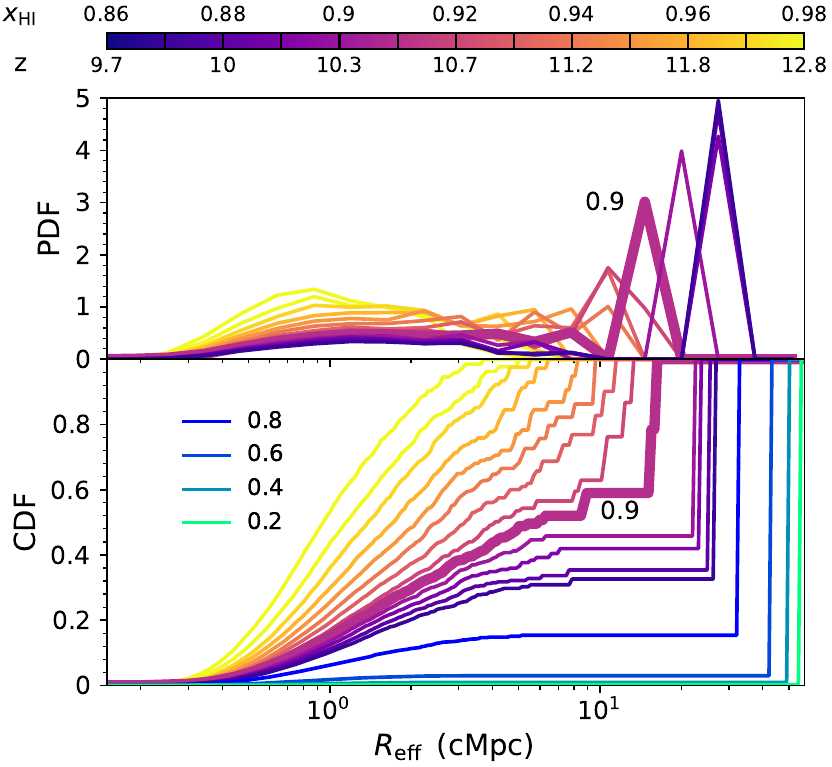}
    \caption{\textbf{Evolution of the Bubble Size Distribution:} Volume-weighted PDFs and CDFs of effective bubble radii $R_\text{eff}$ at different redshifts for the unsmoothed data. The distribution shifts towards larger sizes over time, with a significant reduction in intermediate-sized bubbles ($R_\text{eff} \approx 10\,\text{cMpc}$) after $z \approx 10$ or a global neutral fraction of $0.9$ due to mergers with the largest bubble.
    The redshift values are chosen to bracket the main merger event, after which the main bubble dominates the distribution.}
    \label{fig:zreff}
\end{figure}

In the bottom panel of Figure~\ref{fig:fmain}, we see that the fraction of growth in the largest bubble ($f_\text{main}$) transitions from zero to one around $z \approx 10$. At higher redshifts, the growth of all bubbles dominates over the growth of the main bubble, which is expected since all bubbles start with only one cell. Furthermore, some bubbles at this early time ($z\gtrsim 11$) are larger than the main bubble. However, the difference is insignificant as there are several bubbles of relatively equal size that compete to emerge as the one that happens to be the largest. By $z \approx 10-11$, the $f_\text{main}$ rapidly increases, indicating that the combined percolation of neighbouring bubbles and expansion assigned to the main bubble is already globally important. This is also when the main bubble undergoes its ``main merging event,'' synonymous with its merging stage of growth. During the main merging event, the largest bubbles (and many smaller ones) merge into the main bubble, resulting in a bubble much larger than any others. Since it is much larger than any other bubble, it accounts for nearly all of the ionized volume. Thus, at the tail end of the main merging event at $z \lesssim 9$, the $f_\text{main}$ gradually approaches one as both the largest bubble and the whole box undergo the final stage of rapid expansion. Since the $f_\text{main}$ is so close to unity at $z \lesssim 8$, the expansion of the simulation closely matches the expansion of the largest bubble. Therefore, the largest bubble establishes itself as the main bubble after its main merging event, dominating the simulation beginning at $\approx 10$ per cent global ionization fraction. A key takeaway is that this dominance is established long before the box is reionized, which occurs later around $z \sim 7$.

\subsection{Bubble sizes and morphologies}
\label{subsec:bubblesize}
Understanding the size distribution and morphology of ionized bubbles is essential for interpreting reionization observations. In this section, we analyze the time evolution of bubble volumes (Section~\ref{subsubsec:time_evolution}), and examine the final distribution of bubble sizes at the end of the simulation $z\approx5.5$ (Section~\ref{subsubsec:finaldist}).

\begin{figure}
    \centering
    \includegraphics[width=\columnwidth]{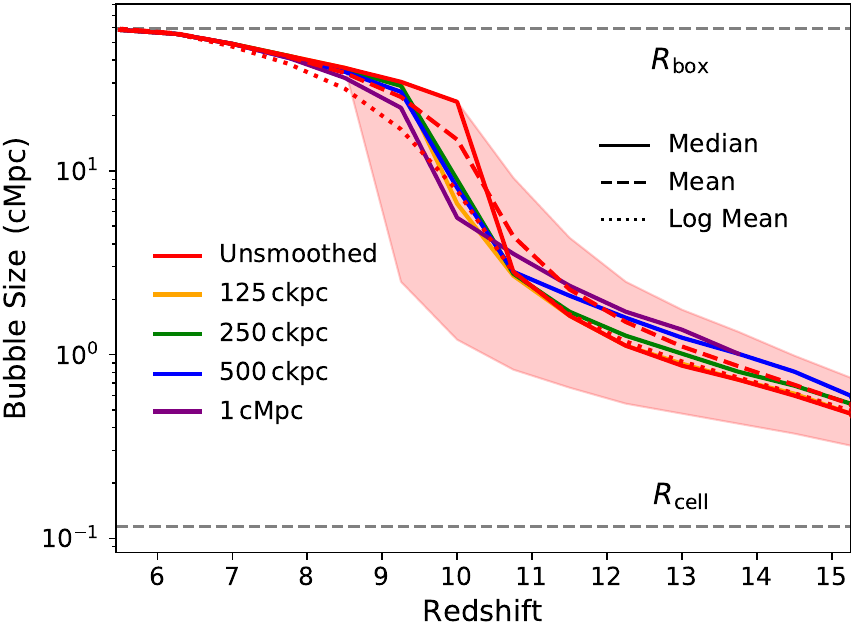}
    \caption{\textbf{Statistical Evolution of Bubble Sizes:} Volume-weighted median, mean, and logarithmic mean effective bubble radii $R_\text{eff}$ as a function of redshift. The dashed lines indicate the spherically-equivalent radii corresponding to one cell ($R_\text{cell}$) and the entire box ($R_\text{box}$). The shaded region represents the $16^\text{th}$ to $84^\text{th}$ percentiles. The statistics reveal the initial expansion, merging, and rapid expansion phases, as all bubbles start off relatively small and grow in a complex percolation process. After the largest bubble goes through its main merging event it becomes much larger than all other bubbles.}
    \label{fig:median}
\end{figure}

\subsubsection{Time evolution of bubble sizes}
\label{subsubsec:time_evolution}
As reionization progresses, ionized bubbles grow and merge, leading to changes in the size distribution over time. From the bubble tree outputs we can calculate the distribution of (spherically-equivalent) effective bubble radii ($R_\text{eff}$) throughout the early EoR. In Figure~\ref{fig:zreff}, we present the volume-weighted probability density function (PDF) and cumulative distribution function (CDF) of $R_\text{eff}$ at various redshifts for the unsmoothed data. The top panel only shows the \HI fraction in the range 0.86--0.98 ($z = 9.7 - 12.8$) because the largest bubble dominates the simulation after this, causing no visible distribution apart from the largest bubble. However, lower \HI fractions are shown in the bottom panel. As the redshift decreases, the original broad distribution of small sizes decreases as a peak of large sizes forms. The peak on the right is from the main bubble, and is sharp because there is only one bubble contributing to that size range. Furthermore, on the bottom panel, starting around $x_{\HI}$ $\approx0.9$, the bubbles of size 10\,cMpc start to disappear. We see this in the flattening of the CDF around this bubble size, with smaller sizes also disappearing from the simulation as the simulation progresses. Thus, once the main bubble finishes its main merging event it has merged with all of the intermediate-sized bubbles ($R_\text{eff} \approx 10\,\text{cMpc}$), and continues to merge with smaller bubbles before they grow beyond a few cMpc. Lastly, around $x_{\HI}$ $\approx 0.6$, the main bubble is much larger than the other bubbles, resulting in smaller bubbles having almost no impact on the CDF, due to the volume weighting of the statistics.

Once the largest bubble goes through its main merging event, the distribution of bubble sizes remains fairly constant. To show this, Figure~\ref{fig:median} displays the evolution of the volume-weighted median, mean, and logarithmic mean bubble sizes as a function of redshift. The shaded region represents the $16^\text{th}$ to $84^\text{th}$ percentiles. The characteristic bubble sizes start off small as the bubbles have not yet had the time to grow significantly. However, after the main bubble's merging event, $z \approx 9-10$ (see Figure~\ref{fig:fmain}), the percentile range collapses to the median, and the three statistics approach each other as they slowly grow, due to the main bubble dominating the box.

\begin{figure}
    \centering
    \includegraphics[width=\columnwidth]{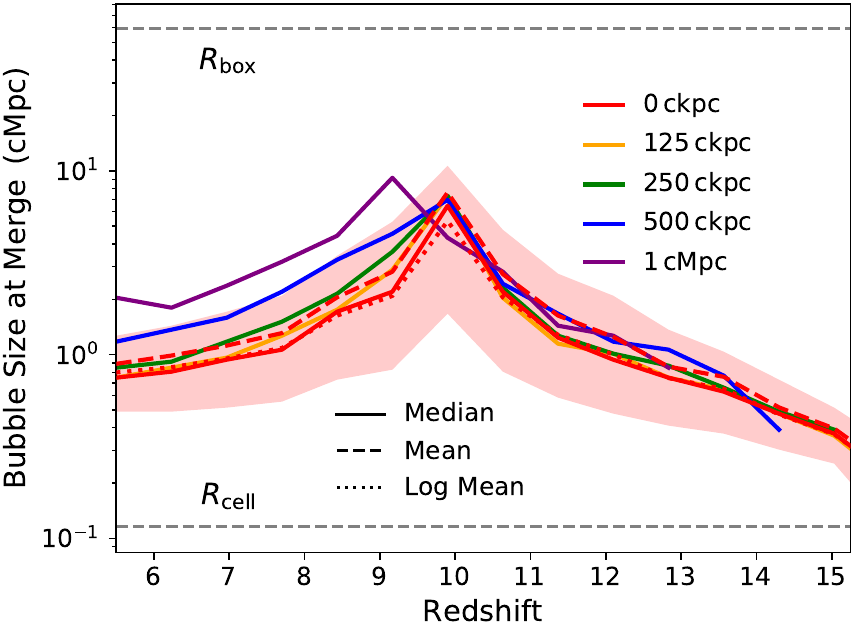}
    \caption{\textbf{Bubble sizes at Merging:} Volume-weighted median, mean, and logarithmic mean effective bubble radii $R_\text{eff}$ prior to merging, plotted against the redshift of merging events. The shaded region represents the $16^\text{th}$ to $84^\text{th}$ percentiles. A peak occurs during the main merging event, before which bubble growth is not prevented, and after which bubble sizes decrease as the largest bubble absorbs smaller ones.}
    \label{fig:rmerge}
\end{figure}.

To investigate the sizes of bubbles at the time of merging, Figure~\ref{fig:rmerge} presents the median, mean, and log mean sizes just before merging, as a function of the redshift at which each merger event takes place. Once again, all statistics are volume-weighted but the partial expansion of unmerged bubbles is not included. We observe a peak in bubble sizes during the onset of the main bubble's merging event, followed by a sharp decrease at the end of the event. At this point, bubble sizes continue to decrease, indicating that the main merging event incorporates all of the intermediate-size bubbles into the largest bubble and prevents any remaining bubbles from reaching a similar size as they merge before growing larger than $R_\text{eff} \approx 1\,\text{cMpc}$.

\begin{figure}
    \centering
    \includegraphics[width=\columnwidth]{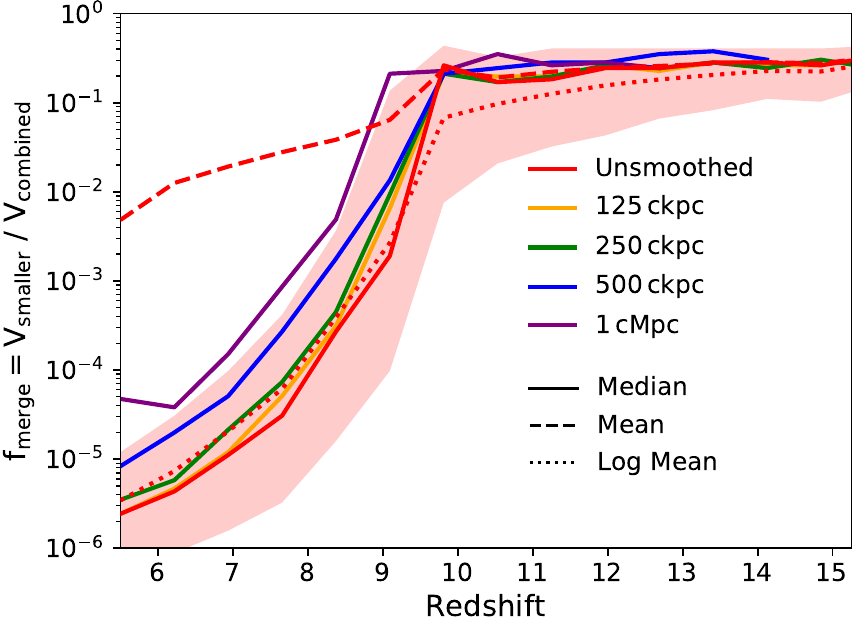}
    \caption{\textbf{Evolution of Merger Ratios:} Statistics for the ratio of smaller-to-combined volumes $f_\text{merge}$ for merger events as a function of the merger redshift. Early on, the bubbles are of equal size, resulting in higher merger ratios. After the main bubble goes through its merging event, it is much larger than the other bubbles, leading to very small values of $f_\text{merge}$. The mean departs from the median trend because some near-equal-volume mergers are still happening with the smaller bubbles.}
    \label{fig:fmergez}
\end{figure}

To explore the volume ratio of mergers throughout the simulation we calculate $f_\text{merge}$, defined for a merger event as the volume of the smaller bubble divided by the combined volume of the merging bubbles. Figure~\ref{fig:fmergez} displays the evolution of $f_\text{merge}$ with the redshift at which the merger event occurs. The statistics are weighted by the volume of the smaller bubble. The fraction starts off high, around $0.2-0.4$, however, around redshift $\approx 9-10$, there is a sharp decrease in $f_\text{merge}$ corresponding to the time at which the main bubble undergoes its merging event.

The high-redshift portion of the plot is flat due to all the bubbles being relatively small and of similar sizes as they merge with each other early on. Then, as the main bubble grows much larger, it dominates the mergers, causing a significant decrease in $f_\text{merge}$. The mean does not decrease as sharply as the median and log mean because smaller bubbles continue to merge with others of similar sizes. Since the mean is not resistant to outliers, it does not decrease as significantly as the other metrics.

\subsubsection{Final Size Distribution}
\label{subsubsec:finaldist}
We now examine the final size distribution of all bubbles across the entire simulation, effectively marginalizing over the redshift evolution. In practice, there are two size distributions of interest corresponding to expansion and merging events. For the first, we assign an effective radius $R_\text{eff}(z_\text{reion})$ to each cell based on the volume of the bubble at the time it is grown into, i.e. its reionization redshift. Since only one bubble grows into each cell, the effective radius for expansion is unique. For the second, we simply adopt the effective radius $R_\text{eff}(z_\text{merge})$ from each merger event based on the volume prior to merging. The merger-based distribution represents a subset of the points included in the expansion-based one, providing a complementary perspective on the various growth phases.

\begin{figure}
    \centering
    \includegraphics[width=\columnwidth]{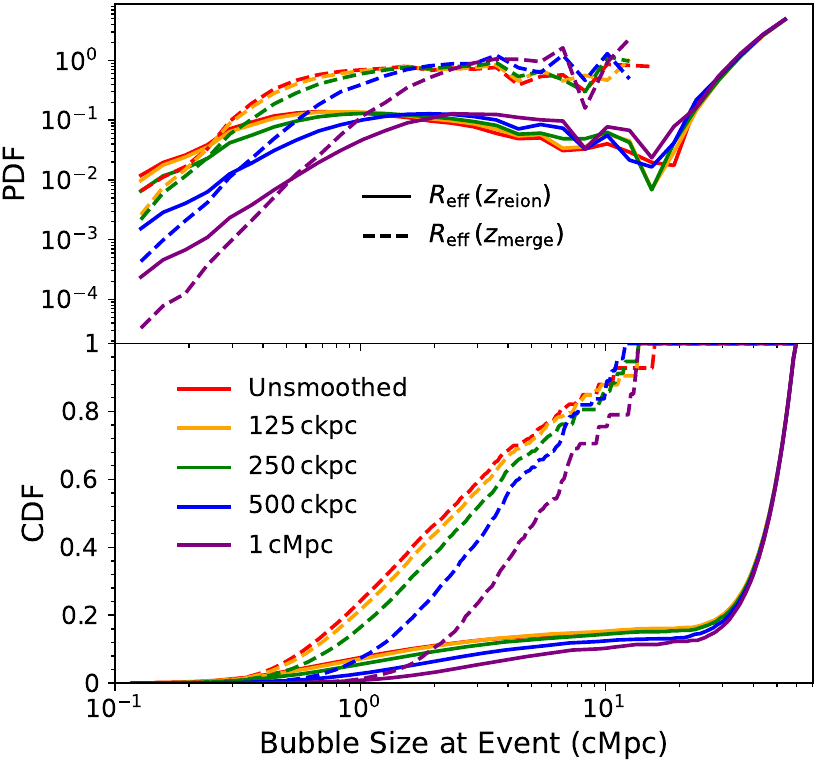}
    \caption{\textbf{Final Distribution of Bubble Sizes:} Volume-weighted PDFs and CDFs of the effective bubble radii $R_\text{eff}$ marginalized over all expansion ($z_\text{reion}$; solid) and merging ($z_\text{merge}$; dashed) events (see the text for more details). For $z_\text{reion}$, we see a broad peak below $R_\text{eff} \lesssim 10\,\text{cMpc}$, showing an abundance of small-to-intermediate bubbles. However, there is a sharp peak at the end, where many cells are being grown into the largest bubble. For $z_\text{merge}$, the curves suddenly end because there is only one bubble larger than that size, the main bubble, which never merges.}
    \label{fig:rfreq}
\end{figure}

Figure~\ref{fig:rfreq} shows the volume-weighted PDF and CDF of the effective bubble radii $R_\text{eff}$ for both expansion ($z_\text{reion}$) and merging ($z_\text{merge}$) events. Figures~\ref{fig:median} and~\ref{fig:rmerge} are the time-resolved versions of the plot. There is a slight peak around $R_\text{eff}(z_\text{reion}) \approx 1\,\text{cMpc}$, indicating an abundance of small bubbles that expanded to this size before merging. Since this radius is relatively small, only a few expansions are required to grow out of this size range. This suggests that many small-to-intermediate bubbles exist, and contribute a non-negligible fraction of the CDF before the emergence of the main bubble. Then, we observe a decrease in the frequency of bubble sizes as the effective radii increase to $\approx 10\,\text{cMpc}$. This is due to the largest bubble absorbing the other large bubble groups during its main merging event (see Figure~\ref{fig:zreff}). This, in effect, allows the largest bubble to quickly grow through these sizes while also preventing other bubble groups from expanding to this size, producing a dip in the PDF. Finally, there is a sharp spike at $R_\text{eff}$, reflecting the increasing dominance of the largest bubble during the final expansion phase.
Given the bubble's large size, individual additions do not significantly alter its radius. Therefore, many cells join bubbles of this volume before their radius changes substantially.
For $R_\text{eff}(z_\text{merge})$, there is a sharp cutoff at $R_\text{eff} \approx 10\,\text{cMpc}$ (approximately $10^{-3}$ of the simulated volume) because no other bubbles are able to grow to that size before merging. There is essentially only one bubble past that range, the main bubble. The PDFs from the two events exhibit similar shapes, albeit with different normalizations as the largest bubble never merges. This impacts the resulting CDFs, which are more linear with the merging statistics and reveals a clear trend of larger sizes at higher smoothing levels.

\begin{figure}
    \centering
    \includegraphics[width=\columnwidth]{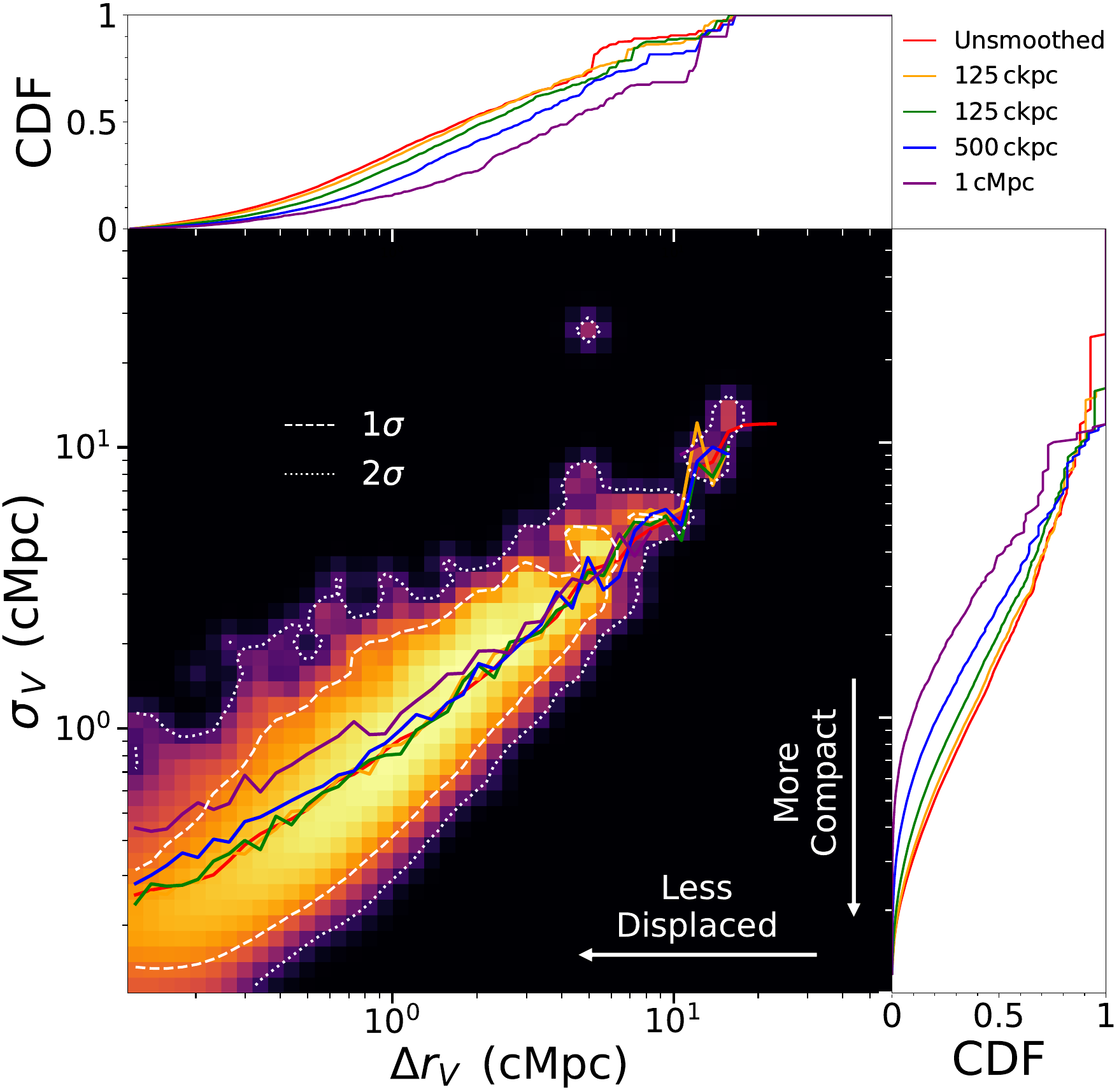}
    \caption{\textbf{Bubble Morphology:} Distribution of bubble centre-of-volume spatial standard deviation $\sigma_V$ versus displacement relative to its starting point $\Delta r_V$, coloured by the relative contribution per dex$^2$ weighted by the volume of the smaller bubble. $1\sigma$ and $2\sigma$ contours and running median curves are included to aid with visual interpretation. The top and side panels show the volume-weighted CDFs for $\Delta r_V$ and $\sigma_V$, respectively. A positive correlation suggests that larger, less compact bubbles have centres of volume farther from their starting points.}
    \label{fig:rcom}
\end{figure}

\begin{figure}
    \centering
    \includegraphics[width=\columnwidth]{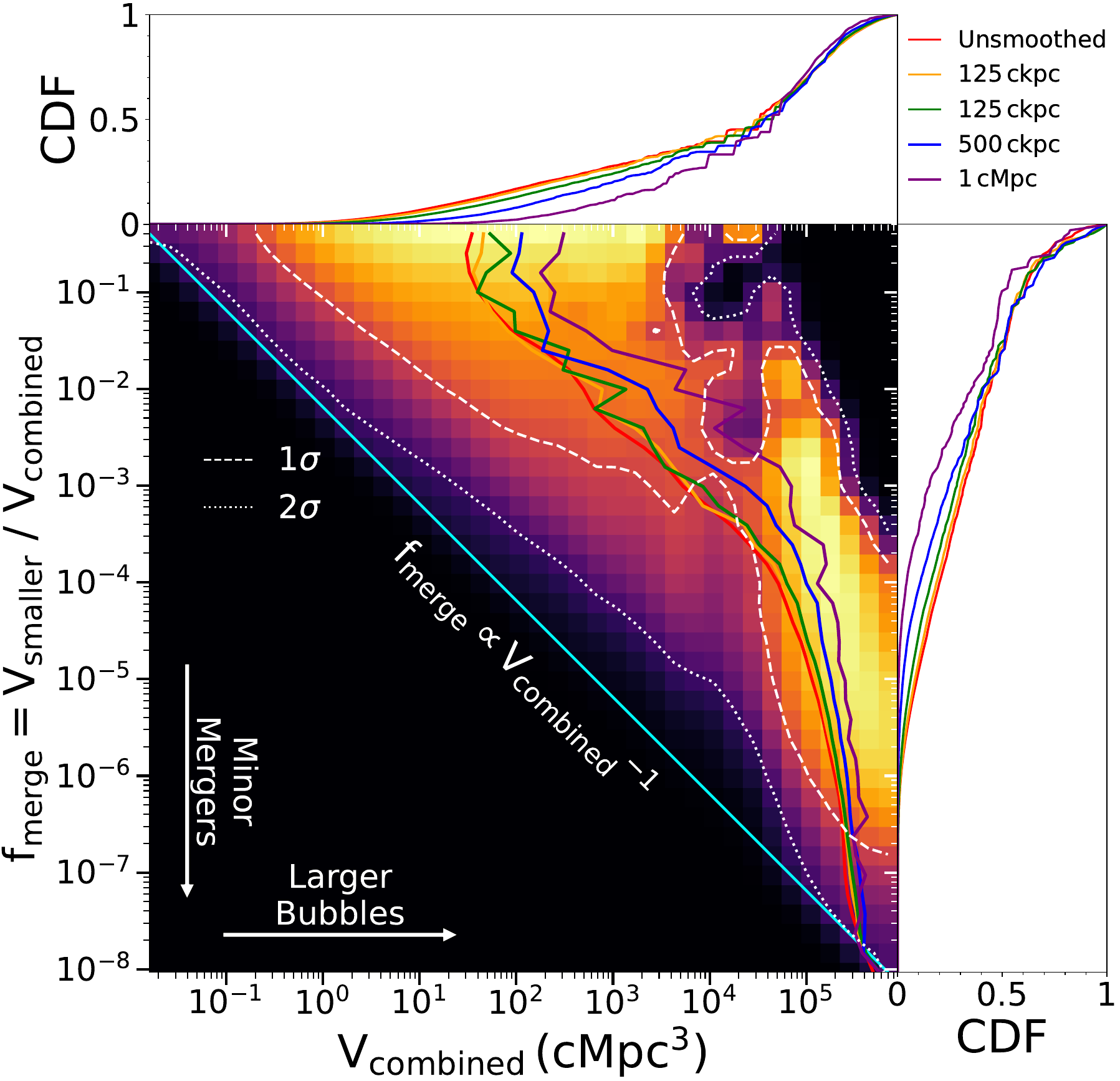}
    \caption{\textbf{Merger Ratios vs. Combined Volumes:} Distribution of bubble merger ratios $f_\text{merge}$ versus the combined volume upon merging $V_\text{combined}$, coloured by the relative contribution per dex$^2$ weighted by the volume of the smaller bubble. $1\sigma$ and $2\sigma$ contours and running median curves are included to aid with visual interpretation. The two peaks correspond to mergers between similar-sized bubbles and mergers where small bubbles are absorbed by much larger ones.}
    \label{fig:fmergevc}
\end{figure}

To assess bubble morphologies, we analyze the displacement of the bubble centres of volume relative to their initial seed points $\Delta r_V$ and their spatial standard deviation $\sigma_V$ as a measure of compactness. These two metrics represent a first look at the bubble's shape at its end state. Figure~\ref{fig:rcom} illustrates the joint distribution of $\Delta r_V$ and $\sigma_V$ and is coloured by the relative contribution per dex$^2$. All statistics are volume-weighted, and the largest bubble is omitted because it encompasses the entire box at its end state. We find a positive correlation between the displacement and spatial extent of bubbles, indicating that larger, more irregular bubbles tend to have centres of volume farther from their initial positions. This is most likely related to the size of the bubble. As the bubble gains more cells, it has a higher opportunity to get further away from the starting cell, and become more irregular. Furthermore, smaller bubbles do have the opportunity to get large enough to get a centre of volume displacement as large as shown in the figure. Therefore, the larger the bubble, the more off-centre it becomes. In addition, we find no significant deviation from the expected correlation of $\sigma_V\sim R_\text{eff}$. In fact, the mild dependence is $\text{d}\log\sigma_V / \text{d}R_\text{eff}\approx0.056$

Similarly, in Figure~\ref{fig:fmergevc}, we plot the merger ratio $f_\text{merge}$ against the combined volume of the merging bubbles $V_\text{combined}$, as defined above. Interestingly, there are two peaks in this distribution. The first one is where the smaller bubbles merge with each other with nearly equal sizes. The second is where the bubbles merge with very unequal sizes. These results are consistent with Figure~\ref{fig:fmergez}, as the first peak is early on in the simulation, and the latter peak is later in the simulation history.

We further characterize the final bubble morphologies upon merging by utilizing the inertial tensor, which has been used as an ionized bubble metric by \citet{Thelie2022},
\begin{equation}
    I_{i,j} = \sum_k (x_{k,i} - x_{\text{ref},i}) \cdot (x_{k,j} - x_{\text{ref},j}) \, .
\end{equation}
Here, $k$ sums over all of the cells within each bubble group, and $i,j$ corresponds to Cartesian coordinates where $i,j\in\{1,2,3\}$. $x_\text{ref}$ denotes the reference point, which we define as the initial starting point of each group. Since we are only interested in the shape, we apply an equal weight to each cell. We characterize the geometrical shape using the eigenvectors and eigenvalues of the inertial tensor. The vectors provide the extension directions of each group, with the highest eigenvalue corresponding to the main elongation axis along which ionization propagates easiest for each bubble. From the eigenvalues, we calculate the triaxiality parameter, defined as follows:
\begin{equation}
\label{eq:triaxiality}
    T = \frac{\lambda_3^2 - \lambda_2^2}{\lambda_3^2 - \lambda_1^2} \, ,
\end{equation}
where $\lambda_1 \leq \lambda_2 \leq \lambda_3$ are the eigenvalues of the inertial tensor. A value of $0 < T < \frac{1}{3}$ indicates the object is oblate with two large dimensions and one small dimension (disk-like). Whenever $\frac{1}{3} < T < \frac{2}{3}$, the object is triaxial such that all three dimensions are significant. When $\frac{2}{3} < T < 1$ the object is prolate with two small and one large dimension (filament-like).

\begin{figure}
    \centering
    \includegraphics[width=\columnwidth]{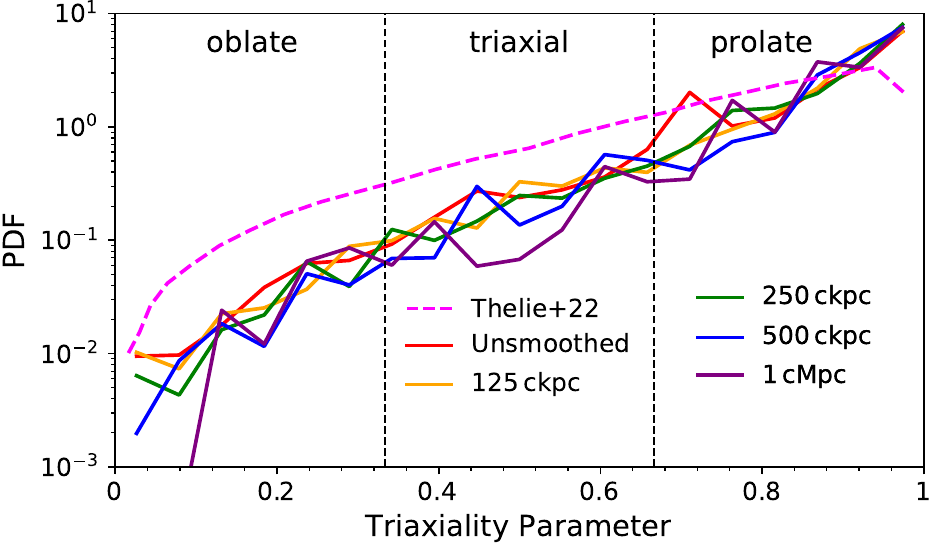}
    \caption{\textbf{Bubble Triaxiality:} Distribution of the triaxiality parameter $T$ for the bubbles, as defined in Eq.~(\ref{eq:triaxiality}). $T$ characterizes the bubble shapes as being oblate (disk-like), triaxial, or prolate (filament-like). We compare our results to a similar calculation by \citet{Thelie2022}. The general linear increase is consistent, indicating a higher prevalence of prolate shapes among bubbles independent of the reionization simulation and the volume segmentation algorithm.}
    \label{fig:triaxiality}
\end{figure}

Figure~\ref{fig:triaxiality} shows the distribution of triaxiality parameters for the bubbles in their final state, which we compare to the results found by \citet{Thelie2022}, which used semi-analytical
\textsc{21cmFAST} and fully numerical \textsc{EMMA} simulations. We emphasize that this is not an apples-to-apples comparison as the volume segmentation algorithms are different. However, the general increasing trend is consistent between the two, as more bubbles are prolate than oblate. However, our curves are steeper and do not exhibit the slight dips at the extremes of the $T$ range (zero and one) found in \citet{Thelie2022}. Also, higher smoothing levels have a smaller number of bubbles initially, so there is larger noise in the binning.

\section{Physics Variation Comparisons}
\label{sec:physics_comparison}
\begin{figure}
    \centering
    \includegraphics[width=\columnwidth]{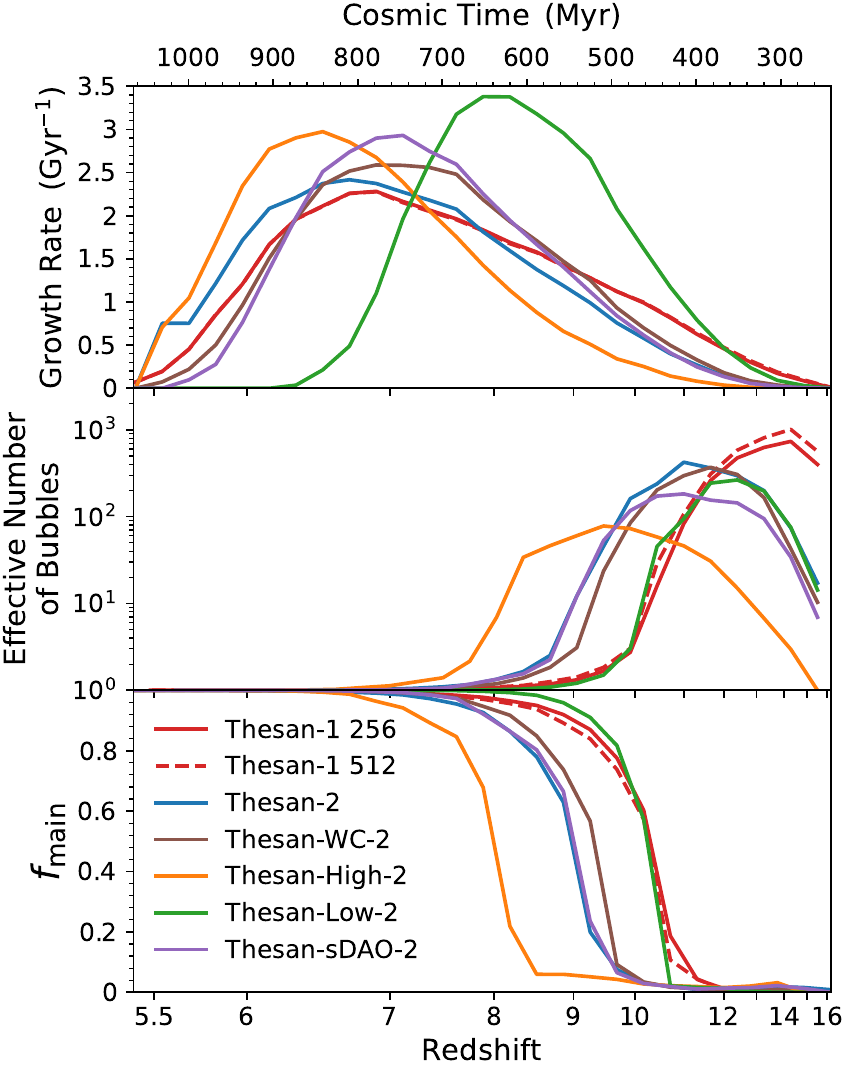}
    \caption{\textbf{Impact of Physics Variations on Bubble Growth:} Similar to Figures~\ref{fig:growthrate}, \ref{fig:neff}, and \ref{fig:fmain} but for the different runs in the \thesan suite. \textit{Top panel}: Growth rate of ionized bubble regions. \textit{Middle panel}: Effective number of actively growing bubbles. \textit{Bottom panel}: Fraction $f_\text{main}$ of cells ionized by the largest bubble. Simulations cluster into three groupings based on their differences, mainly in the overall timing of reionization as being earlier or later with a longer or shorter duration. The \thesanone results at $512^3$ and $256^3$ resolutions show a minimal impact from resolution differences.}
    \label{fig:thesan_plots}
\end{figure}

While our previous analyses focused on the fiducial \thesanone simulation, the \thesan suite includes several other simulations that explore different physical parameters. In this section, we examine how these variations affect the growth and size distribution of ionized bubbles. Specifically, we compare six simulations: \thesanone is the fiducial model, \thesantwo is the same but with two times lower spatial resolution, \thesanwc adjusts the birth cloud escape fraction to match the \,\thesanone reionization history, \thesanhigh only allows escape from higher-mass halos with $M_\text{halo} > 10^{10}\,\Msun$, \thesanlow only allows escape from lower-mass halos with $M_\text{halo} < 10^{10}\,\Msun$, and \thesansdao employs an alternative dark matter model \citep[see][for details]{Kannan2022a, Garaldi2023}. For this comparison, we use a smoothing of 125\,ckpc, which provides some smoothing without significantly altering the morphology of bubbles. We adopt a grid resolution of $256^3$ cells to reduce computational and storage requirements for these high-level explorations. Tests with \thesanone show that results at this resolution are consistent with the main ones with $512^3$ cells. Briefly, Figure~\ref{fig:thesan_plots} presents the global growth rate, effective number of bubbles, and the fraction $f_\text{main}$ of cells ionized by the largest bubble for each simulation.

In the top panel of Figure~\ref{fig:thesan_plots}, we find some significant variations in the growth rate compared to the fiducial \thesanone simulation. Specifically, \thesanlow exhibits a much earlier and higher peak in the growth rate, indicating that lower-mass halos contribute more strongly to early reionization when they are the primary sources of ionizing photons. \thesanhigh shows a delayed peak, reflecting the reliance on higher-mass halos, which are more abundant at later times. The remaining \thesantwo, \thesanwc, and \thesansdao have more similar growth rates. These simulations show a delayed start compared to \thesanone, which remains as a delayed end for \thesantwo, but as time progresses, they begin to deviate from each other with the peaks of \thesanwc and \thesansdao aligning with \thesanone.

The middle panel shows the effective number of actively growing bubbles. Here we see that \thesanone and \thesanlow start with significantly more bubbles actively growing than the others, with \thesanone having the earliest peak, followed by \thesanlow, which are both dominated by lower-mass haloes earlier on. In contrast, \thesanhigh reaches its peak in the effective number of bubbles at later times, consistent with the evolving abundance of higher-mass haloes. Additionally, the other simulations (\thesantwo, \thesanwc, and \thesansdao) have a more gradual increase in the number of bubbles, reflecting intermediate behaviours. Interestingly, there is no significant difference between the results from the two resolutions (256 and 512 cells).

The bottom panel examines $f_\text{main}$, the fraction of cells expanded into by the main bubble relative to the total number of cells expanded in the simulation. The \thesanone and \thesanlow simulations show the earliest rise in $f_\text{main}$. Although one might expect \thesanlow to differ significantly from \thesanone in the growth of the largest bubble (given the differences in the growth rate and effective number of bubbles), the bottom panel shows that they are relatively similar in $f_\text{main}$. This is because, in both simulations, the largest bubble begins as a single cell and initially grows slowly. The largest bubble becomes dominant sooner in these simulations, because lower-mass sources lead to an earlier reionization and the smaller mean bubble separations change the morphology of the percolation process. On the other hand, \thesanhigh has the latest timing for the main bubble merging as it needs to wait for a critical number of higher-mass haloes. The remaining simulations group together between these two behaviours, with $f_\text{main}$ rising at intermediate times.

\begin{figure}
    \centering
    \includegraphics[width=\columnwidth]{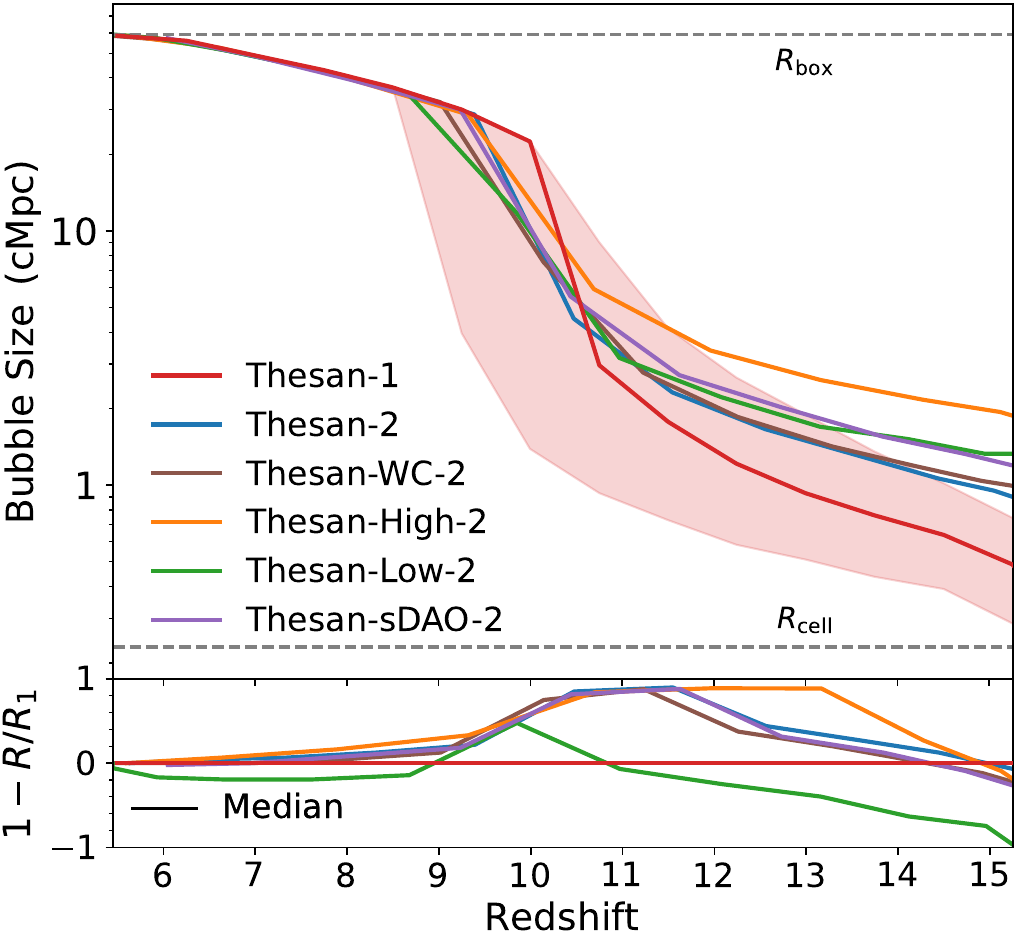}
    \caption{\textbf{Impact of Physics Variations on Bubble Sizes:} \textit{Top panel}: Volume-weighted median bubble size as a function of redshift for different \thesan simulations. The redshift has been corrected for the different reionization histories by the following mapping: $z(\text{\textsc{thesan-x}}) \rightarrow x_\HI(\text{\textsc{thesan-x}}) \rightarrow x_\HI(\text{\textsc{thesan-1}}) \rightarrow z(\text{\textsc{thesan-1}})$. The shaded region indicates the $16^\text{th}$ to $84^\text{th}$ percentiles for \thesanone. \textit{Bottom panel}: Differences in median bubble sizes relative to \thesanone. Variations in physical parameters lead to differences in the timing and extent of bubble growth and merging events.}
    \label{fig:thesan_medians}
\end{figure}

We further explore the impact of physical parameters on the distribution of bubble sizes. Figure~\ref{fig:thesan_medians} displays the volume-weighted median bubble size as a function of redshift for each simulation. The shaded region represents the $16^\text{th}$ and $84^\text{th}$ percentiles for \thesanone. Consistent with previous observations, we identify three groups of simulations: \thesanone and \thesanlow, where the main bubble grows in size much sooner due to the earlier ionizing output from lower-mass haloes; \thesanhigh, which experiences significant growth later in the simulation due to the delayed activity of higher-mass haloes; and the rest of the simulations, which fall between these two extremes. The bottom panel shows the differences in median bubble size relative to the base \thesanone simulation. The observed variation can be explained by the timing and duration of the buildup, rapid growth, and aftermath of the main merging event of \thesanone. Although not shown, we do not find qualitatively different conclusions from examining other statistics such as $f_\text{merge}$. Still, these results demonstrate that varying model parameters can influence the efficiency of bubble growth and merging during reionization, acting as a sensitive probe in theoretical studies seeking to discriminate between ionizing source scenarios.

\section{Conclusions}
\label{sec:conclusion}
Analyzing ionized bubbles offers valuable insights into the large-scale processes of the EoR. While various techniques have been developed to identify and characterize these bubbles, we utilized the reionization redshift (\zreion) to study and describe their growth and morphology. We developed a spatiotemporal algorithm based on \zreion data from the \thesan simulations to examine bubble growth rates across different smoothing scales, the final bubble size distribution, the time evolution of bubble sizes, and the impact of physics model variations.

Our analysis of \zreion and its connection to ionized bubbles has led to several insights into the progression of reionization in the \thesan simulations. We summarize our main findings as follows:
\begin{enumerate}
    \item There are three distinct stages of growth for bubbles during the EoR. First, an initial period of isolated expansion while the bubbles are far away from each other, from $z \approx 11-16$. Once the bubbles are large enough to come into contact with each other, the second phase of accelerated growth through merging begins, from $z \approx 9-11$. After most mergers are resolved, a single dominant bubble expands rapidly, ionizing the remaining neutral regions, from $z \approx 9-5.5$. Additionally, the lower smoothing levels exhibit higher growth rates during the first two phases, whereas higher smoothing levels show higher rates during the final phase.
    \item At high redshifts, numerous bubbles are actively growing, with most having radii around $\approx0.5$\,cMpc (marginally resolved on uniform grids). However, once the largest bubble percolates around $z \approx9-10$, it becomes significantly larger than other bubbles and dominates the simulation, with statistical measures converging toward those of the main bubble.
    \item Bubble sizes around $\approx 10$\,cMpc are notably underrepresented in the size distribution. This occurs because, during the main merging event, the largest bubble absorbs all other large bubble groups, effectively bypassing the $\approx 10$\,cMpc size range. Furthermore, since the smaller bubbles never get large enough to fill in this size before merging, this gap in the size distribution persists.
    \item We investigated the impact of pre-smoothing the volume-weighted ionized density field and found that our conclusions remain robust unless the smoothing scale becomes large enough to alter the morphology of bubbles. We recommend a smoothing scale of approximately 125\,ckpc, as larger scales (e.g., 1\,cMpc) can significantly affect the timing of bubble growth, particularly delaying the growth of small bubbles at early times.
    \item Variations in physical parameters across different \thesan simulations affect the timing of the three growth stages. This results in three clusters of simulations that share similar timing patterns. \thesanlow and \thesanone exhibit the earliest growth due to the significant contribution of lower-mass haloes to reionization. \thesanhigh shows the latest onset of growth due to the delayed contribution of higher-mass haloes. Finally, \thesantwo, \thesanwc, and \thesansdao fall between these two extremes. Overall, the model dependence is relatively weak once the reionization history is considered, as the main bubble is already established by the time reionization reaches 10 per cent completion.
\end{enumerate}

These findings provide a unique perspective on the evolution of ionized bubbles during the EoR, including the hierarchical merging process. For future work, this study can be extended by connecting the identified bubble groups to the properties of galaxies within them. By properly associating ionizing sources with their corresponding bubbles, we can track and analyze the local ionizing radiation budget and escape mechanisms. It would also be interesting to employ percolation theory to characterize the geometry of large-scale reionization environments. An essential aspect of cosmic web geometry involves the connectivity of its components, such as the clustered hierarchy of filaments and voids. Percolation methods have been used to compare the geometrical properties of the real cosmic web with the simulated dark matter web \citep{Einasto2018}, and to examine continuum percolation statistics for both high-resolution dark matter particle distributions and galaxy mock catalogues from a semi-analytic galaxy formation model \citep{Regos2024}. The insights and discrimination percolation has on the formation of large-scale structures can be applied to the formation and growth of ionized bubbles during the EoR. Notably, the sudden transition where one cluster dominates all others can be compared to the main bubble's growth, revealing further insights into the relationship between galaxy formation and the reionization process.

\section*{Acknowledgements}

NJ is grateful for the physics NSF REU program hosted at UT Dallas and the program's coordinators Lindsay King and Mike Kesden. He thanks Kevin Lorinc and Shrimoy Satpathy for their companionship and support throughout the REU program. The bubble calculations in this paper were performed using the Center for Research Computing at the University of Notre Dame and Lonestar6 at the Texas Advanced Computing Center (TACC). RK acknowledges support of the Natural Sciences and Engineering Research Council of Canada (NSERC) through a Discovery Grant and a Discovery Launch Supplement (funding reference numbers RGPIN-2024-06222 and DGECR-2024-00144) and the support of the York University Global Research Excellence Initiative.
MV acknowledges support through the National Aeronautics and Space Administration (NASA) Astrophysics Theory Program (ATP) 19-ATP19-0019, 19-ATP19-0020, 19-ATP19-0167, and the National Science Foundation (NSF) grants AST-1814053, AST-1814259, AST-1909831, AST-2007355, and AST-2107724.


\section*{Data Availability}

All data produced within the \thesan project are fully and openly available at \url{https://thesan-project.com}, including extensive documentation and usage examples \citep{Garaldi2023}. We invite inquiries and collaboration requests from the community.
Furthermore, a version of our C++ bubble tree algorithm is found at \url{https://github.com/NJamieson22/Bubble-Merger-Tree}.



\bibliographystyle{mnras}
\bibliography{biblio} 




\appendix

\section{Resolution Comparison}
\label{Appendix:Resolution Comparison}
In this appendix, we examine the effects of different grid resolutions on the growth and distribution of ionized bubbles throughout the simulation. We consider three different resolutions: $512^3$, $256^3$, and $128^3$, all with a smoothing scale of 125\,ckpc.

\begin{figure}
    \centering
    \includegraphics[width=\columnwidth]{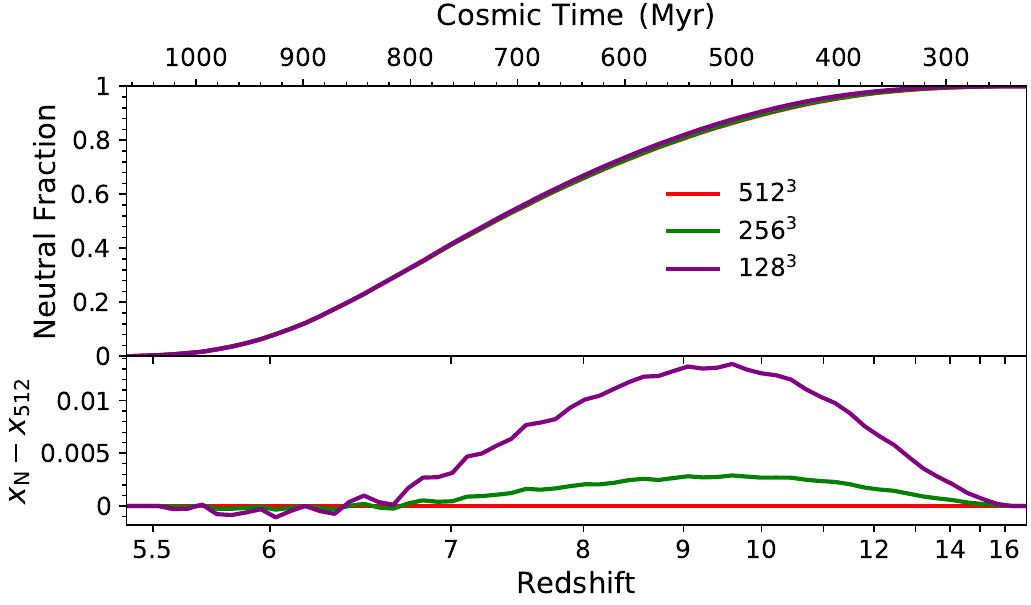}
    \caption{\textbf{Impact of Resolution on Reionization History:} The neutral fraction as a function of redshift for different grid resolutions. As observed, the resolution has very little impact on the growth evolution as determined by the \zreion-inferred persistent reionization history.}
    \label{fig:resolution_hi}
\end{figure}

\begin{figure}
    \centering
    \includegraphics[width=\columnwidth]{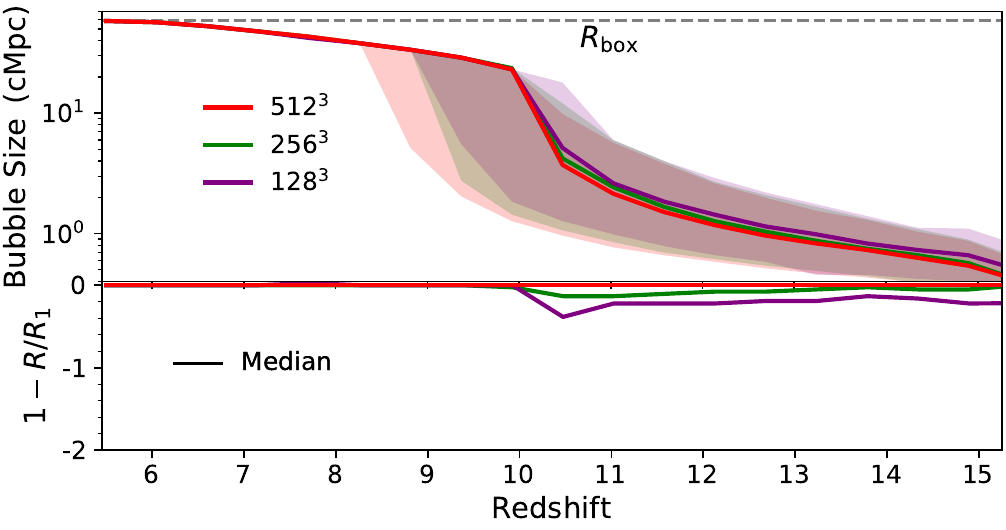}
    \caption{\textbf{Impact of Resolution on Bubble Sizes:} Volume-weighted median bubble sizes as a function of redshift for different resolutions. The sizes are similar but there is a tendency for slightly larger bubbles at early times and the main bubble undergoes its primary merging event earlier in the lower resolutions.}
    \label{fig:resolution_sizes}
\end{figure}

Figure~\ref{fig:resolution_hi} shows the impact of resolution on the reionization history. As observed, changes in resolution have very little impact on the actual growth of bubbles, at the sub-percent level. Next, Figure~\ref{fig:resolution_sizes} shows the volume-weighted median bubble sizes as a function of redshift. While the overall sizes are similar across resolutions, the timing of the main bubbles' merging event occurs slightly earlier for coarse-grained data compared to the higher-resolution ones.

We do not address the convergence of the bubble tree algorithm with box size, as larger boxes are not available in the suite. However, we expect the qualitative results to be the same even though the largest bubble can in principle continue to grow without limit.

\section{Excluding Corners as Neighbours}
\label{Appendix:Corners as Neighbours}
In our bubble tree algorithm, we typically consider corner cells as neighbours, resulting in each cell having 26 adjacent neighbours in three dimensions (including face, edge, and corner neighbours). In this appendix, we explore the effect of excluding corner neighbours, effectively reducing the adjacency to only the six face neighbours per cell. We use a smoothing scale of 125\,ckpc and a grid resolution of 512 for this comparison.

\begin{figure}
    \centering
    \includegraphics[width=\columnwidth]{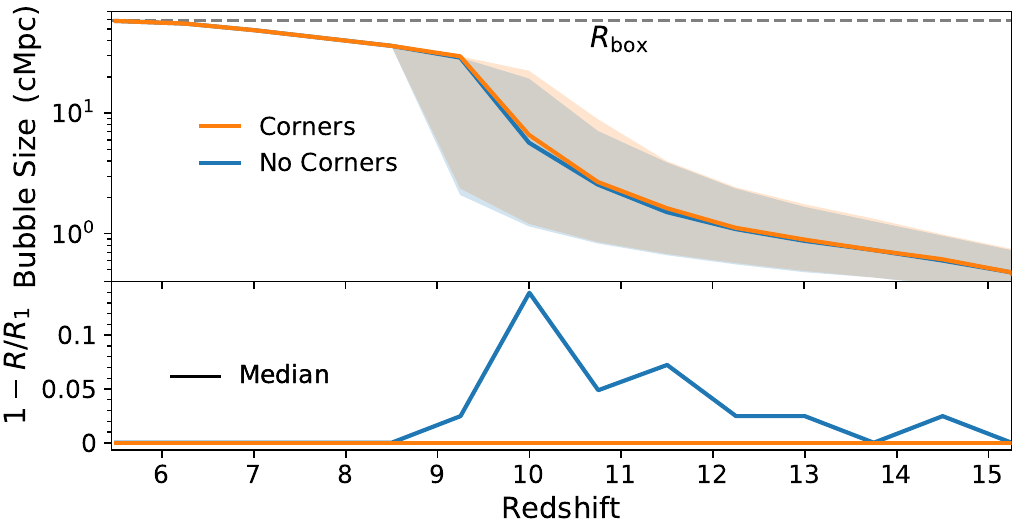}
    \caption{\textbf{Impact of Neighbour Definitions on Bubble Sizes:} Volume-weighted median bubble sizes as a function of redshift, comparing cases where corner neighbours are included (26 neighbours per cell) and excluded (6 neighbours per cell). The resulting curves are nearly identical, indicating no significant difference between these choices.}
    \label{fig:corners_bubblesizes}
\end{figure}

Figure~\ref{fig:corners_bubblesizes} plots the volume-weighted median bubble sizes as a function of redshift for the two scenarios where neighbouring corners are either included or excluded. The results again show minimal difference between the two cases, as the curves are nearly identical. This indicates that either choice is self-consistent and does not substantially impact the overall analysis.

\section{Physics Variation affect on Final Distribution of Bubble Sizes}
\label{Appendix:Thesan Final Sizes}
\begin{figure}
    \centering
    \includegraphics[width=\columnwidth]{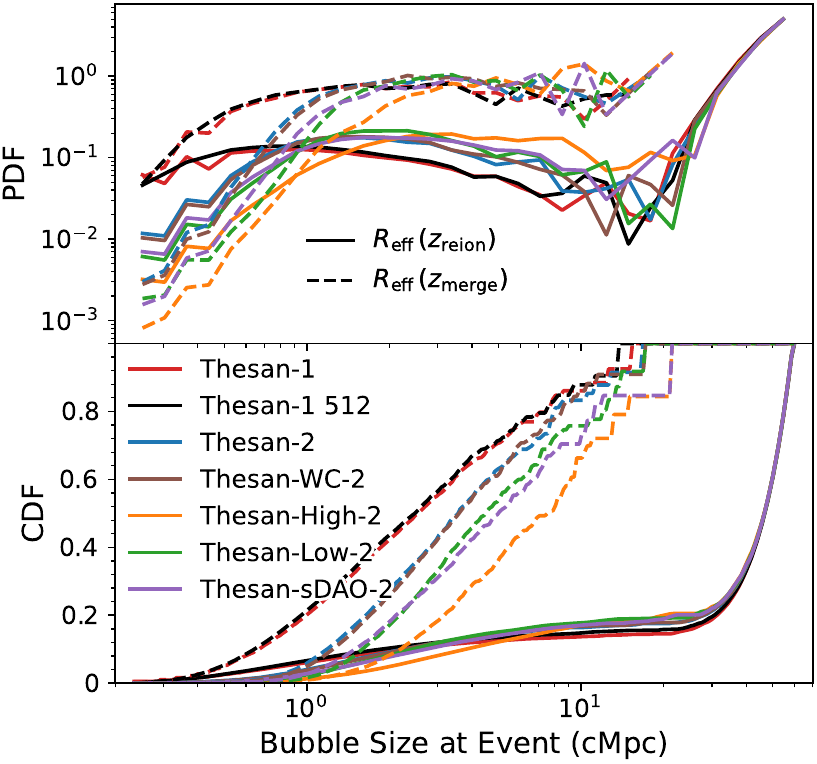}
    \caption{\textbf{Physics Variation on Final Distribution of Bubble Sizes:} Similar to~\ref{fig:rfreq}, we examine the effect of the variation of physical parameters on the volume-weighted PDFs and CDFs of the effective bubble radii $R_\text{eff}$ marginalized over all expansion ($z_\text{reion}$; solid) and merging ($z_\text{merge}$; dashed) events (see the text for more details). The physics variations change the relative frequency of small- and intermediate-sized bubbles.}
    \label{fig:thesan_final_size}
\end{figure}

Similar to Figure~\ref{fig:rfreq}, we examine the effects of the physics variations of the six different \thesan simulations on the final distribution of bubble sizes. We display these results in Figure~\ref{fig:thesan_final_size}.

\bsp	
\label{lastpage}
\end{document}